\documentclass[twocolumn,english,aps,prl,superscriptaddress,nofootinbib,longbibliography]{revtex4-2}
\usepackage[latin9]{inputenc}
\usepackage{xcolor}
\usepackage{babel}
\usepackage{amsmath}
\usepackage{amssymb}
\usepackage{graphicx}
\usepackage[pdfusetitle,
 bookmarks=true,bookmarksnumbered=false,bookmarksopen=false,
 breaklinks=false,pdfborder={0 0 1},backref=false,colorlinks=true]
 {hyperref}
\hypersetup{
 urlcolor=blue, citecolor=blue, hyperfootnotes=blue, linkcolor=blue}

\makeatletter
\usepackage{mathrsfs}
\usepackage{epsfig}
\usepackage{amsfonts}
\usepackage[figuresright]{rotating}
\usepackage{dcolumn}
\usepackage{bm}
\usepackage{color}

\makeatother

\begin{document}
\title{Collectively pair-driven-dissipative bosonic arrays:\\
exotic and self-oscillatory condensates}

\author{Yinan Chen}
\affiliation{Department of Physics, California Institute of Technology, Pasadena,
CA 91125, USA}
\affiliation{Wilczek Quantum Center, School of Physics and Astronomy, Shanghai
Jiao Tong University, Shanghai 200240, China}

\author{Carlos Navarrete-Benlloch}
\email{carlos.navarrete@uv.es}
\affiliation{Departament d'Òptica i Optometria i Ciències de la Visió, Universitat
de València, Av. Vicent Andrés Estellés 19, 46100 Burjassot, Spain}
\affiliation{Wilczek Quantum Center, School of Physics and Astronomy, Shanghai
Jiao Tong University, Shanghai 200240, China}

\begin{abstract}
Modern quantum platforms such as superconducting circuits offer exciting
opportunities for the experimental exploration of driven-dissipative
many-body systems in unconventional regimes. One such regime arises
in bosonic systems, where driving and dissipation can nowadays be
engineered through pairs of excitations, rather than through conventional
single-excitation or linear processes. These platforms also enable
collective rather than local pair loss, so that excitations emitted
into the environment originate from a coherent superposition of lattice
sites rather than any specific site. In this work, we analyze the
superfluid phases accessible to bosonic arrays subject to these novel
mechanisms, which are more characteristic of quantum optics, and show
that they lead to remarkable spatiotemporal properties beyond the
traditional scope of pattern formation in either condensed-matter
systems or nonlinear optics alone. In particular, we show that, even
in the presence of residual local loss, the system is stabilized into
an exotic state in which bosons condense along the modes of a closed
manifold in Fourier space. A weak bias drive controls the population
distribution across these modes, providing access to a wealth of patterns:
from periodic and quasiperiodic structures with tunable spatial wavelengths
to condensates that uniformly populate the closed Fourier manifold.
Furthermore, when residual local linear dissipation is balanced by
pumping, new constants of motion emerge that can force the superfluid
to oscillate in time, through a mechanism analogous to that underlying
recently discovered superfluid time crystals. Finally, we propose
a specific experimental implementation for exploring this rich and
unusual spatiotemporal superfluid behavior.
\end{abstract}
\maketitle

\section{Introduction}

In the last couple of decades we have been able to explore quantum
many-body phenomena with a level of control never thought accessible
before. This has been possible thanks to the development of many clean
and controllable experimental platforms that act as so-called quantum
simulators \cite{Cirac12,Bloch12,Blatt12,AspuruGuzik12,Houck12,Altman21},
such as trapped ultracold atoms \cite{Bloch08,Jaksch05,Dutta15}.
These are well isolated devices that essentially behave as closed
quantum systems where a variety of many-body bosonic \cite{Aidelsburger19,Aidelsburger20},
fermionic \cite{Aidelsburger21a,Bloch20,Bloch21,Gross16,Gross17,Hemmerich21,Esslinger15},
or spin Hamiltonians \cite{Gross17spin,Gross13,Kuhr13,Lukin21,Ketterle20}
can be engineered. In combination with very flexible measurement techniques
giving access to a wide range of observables, these systems have allowed
the observation of many physical phenomena ranging from the already
classic insulator-superfluid phase transition of the Bose-Hubbard
model \cite{Bloch02}, to supersolidity \cite{Esslinger17a,Ketterle17,Esslinger17},
topological effects \cite{Gross17,Bloch21,Bloch16topo,Ketterle15topo,Lukin21topo},
or many-body localization \cite{Bloch15mbl,Gross16mbl,Schneider17mbl,Bloch19mblRMP,Aidelsburger19mbl}
and other routes to the breaking of thermalization \cite{Aidelsburger21mblbutno}.
Coupling these systems to the optical modes of a laser-driven lossy
cavity \cite{Vuletic03atomcav,Esslinger13atomcavRMP,Esslinger13atomcav,Hemmerich15atomcav2,Hemmerich15atomcav,Piazza15atomcav,Esslinger19atomcav,Esslinger21atomcav,Esslinger21atomcav2,Skulte24atomcav}
has granted access the domain of driven-dissipative many-body physics
\cite{Diehl16}. These scenarios are intrinsically out of thermodynamic
equilibrium, and the competition between driving, dissipation, and
interactions can stabilize the system into states that can be inaccessible
to closed systems, for example \textcolor{black}{continuous time crystals
\cite{Wilczek12}, as has been theoretically predicted \cite{Iemini18,Tucker18,Buca19a,Buca19b,Kessler19}
and experimentally explored \cite{Kongkhambut22,Kongkhambut24,Chen23,Greilich24,Wu24,Carraro24,Liu25,Jiao25,Huang25}}.
This so-called dissipative-state preparation \cite{Verstraete09}
has provided further motivation for the development of controlled
driven-dissipative experimental platforms. In the bosonic realm which
occupies our current work, perhaps the best explored platforms are
exciton polaritons in semicondutor microcavities \cite{Carusotto13rev,Giacobino20rev,JaqBloch20rev},
among other photonic platforms \cite{Carusotto21rev,Carusotto19rev},
where spontaneous spatial coherence in the form of a wide variety
of patterns, as well topological phenomena has been reported by now.

In recent years, another experimental platform is taking over as a
leading candidate for the implementation of driven-dissipative bosonic
many-body scenarios \cite{Schuster19,Koch13Rev,Houck12}, the so-called
superconducting circuits \cite{Blais21,Krantz19,Gu17rev}. These
solid-state systems are extremely flexible not only in terms of geometric
design, but also in terms of the type of processes that one can engineer
on them. Currently \cite{AbuGhanem25}, 2D arrays with up to about
100 bosonic modes have been developed and used to explore quantum
advantage \cite{GoogleQAI25,Kim23,Supremacy19,Wu21strongadvantage,zhu2021quantum},
complex-state preparation \cite{Mi24,Morvan22}, discrete time crystals
\cite{Mi22TC,zhang22}, localization and thermalization \cite{GoogleQAI26_localization,Andersen25,Chen21therm,Gong21loc,Ye19loc,Xu18loc},
as well as quantum walks and topological phenomena \cite{Mi22,satzinger21,Gong21,Schuster18,Flurin17}.
Here, the intrinsic energies of the bosonic modes are on the microwave
domain, where we are able to synthesize coherent drives with arbitrary
spectral profiles. In addition, the tunneling rates between the modes
can be controlled in real time and can be made complex (artificial
gauge fields). Moreover, interactions between the bosonic modes can
be engineered via the four-wave mixing occurring at the Josephson
junctions that are naturally integrated in these platforms, which
also allow for the implementation of pair driving and dissipation
\cite{Leghtas15,Touzard18,Markovic18,Jiang21rev,Leghtas20,Leghtas23,Leghtas24}.
These are well-known processes in quantum optics, but are quite unconventional
in the many-body models traditionally associated to condensed-matter
physics. \textcolor{black}{Upgrading many-body driven-dissipative models
with these processes can facilitate access to complex quantum states
such as multimode cats \cite{Zapletal22} or exotic superfluid patterns
\cite{WangCaiNB20}, including supersolids \cite{Grigoriou24}.}

\begin{figure*}
\includegraphics[width=0.8\textwidth]{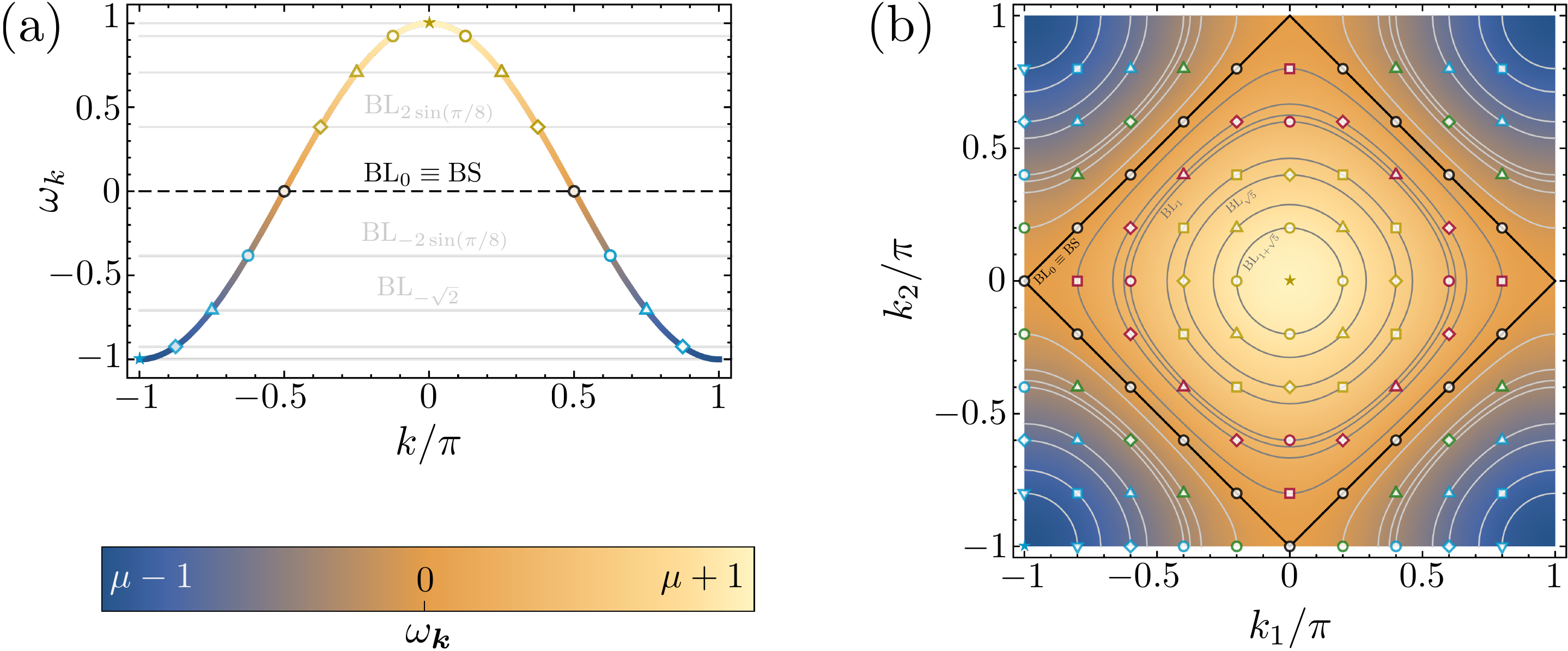} \caption{Dispersion relation $\omega_{\boldsymbol{k}}$ in 1D (a) and 2D (b),
that is, $d=1$ and $2$, respectively. We have chosen $\mu=0$ with
$L=16$ for the 1D case (a) and $L=10$ for 2D (b). We denote by $k$
and $\boldsymbol{k}=(k_{1},k_{2})$ the wave vectors in 1D and 2D,
respectively. We also show the legend for the color scale denoting
the values of $\omega_{\boldsymbol{k}}$, which range between $\mu-1$
(darkest blue) and $\mu+1$ (lighter yellow). Markers of the same
type and color correspond to wave vectors $\boldsymbol{k}$ lying
at the same \textit{Bose level} $\mathrm{BL}_{\beta}$, that is, $\omega_{\boldsymbol{k}}=\beta$.
The Bose level $\mathrm{BL}_{0}$ plays a central role in our work,
since it contains the most divergent modes, which eventually form
the bulk of the asymptotic state; we dub it the \textit{Bose surface}
or $\mathrm{BS}$ \cite{WangCaiNB20}, and highlight it thick black.
Note that the Bose levels $\mathrm{BL}_{\mu\pm2d}$ contain a single
wave vector in all dimensions, for which we have used a star marker.
The rest of Bose levels contain multiple wave vectors, only two in
1D, but many in 2D, eventually forming a closed dense curve in the
thermodynamic limit $L\rightarrow\infty$, represented by the contours
in (b) for the specific values of $\beta$ appearing for $L=10$.}
\label{Fig-Dispersions}
\end{figure*}

This toolbox makes superconducting-circuit arrays a theoretician's
dream for the implementation of interesting and unconventional models
beyond traditional condensed-matter systems. With this motivation,
in this work we study the superfluid phases that emerge in a bosonic
array in the presence of pair driving and the corresponding pair loss,
showing that these lead to incredibly rich spatiotemporal phenomena.
In particular, we consider (and propose a generic implementation for)
the case in which pair loss occurs through a collective channel, i.e.,
the information about which array site the pair came from is washed
off before decaying into the environment. In such case, even in the
presence of additional linear local loss, we show that it is possible
to stabilize the superfluid into an exotic one with bosons condensed
on the modes of a closed manifold in Fourier space (which we also
recently predicted to appear with local pair loss \cite{WangCaiNB20},
but only when the chemical potential and the linear decay are fine-tuned
to zero exactly, which might be too demanding under realistic experimental
conditions). Moreover, at difference with \cite{WangCaiNB20}, the
distribution of population along the ring can be controlled by biasing
the system via the initial conditions or a weak external drive. This
opens the possibility of stabilizing a plethora of patterns, from
periodic and quasi-periodic ones with tunable spatial wavelength,
to homogeneously-populated closed Fourier manifolds. The latter have
been conjectured to play a fundamental role in several open problems
in condensed-matter physics such as high-$T_{\text{c}}$ superconductivity
\cite{Jiang2019}, frustrated magnetism \cite{Sedrakyan2015}, and
interacting problems with spin-orbit coupling \cite{Wu2011,Gopalakrishnan2011}.
Our proposal then opens the way to the systematic study of such broad
type of superfluid patterns under controlled experimental conditions,\textcolor{black}{{}
in the spirit of recent works aimed also at tailoring the condensate's
wave function \cite{Kosior24,Petiziol24}}.

In addition, when the local linear loss is balanced by incoherent
pumping \cite{CNB14,CNB23,Marthaler11prl,Grajcar08,Astafiev07nat},
we show that new constants of motion emerge in the system, which lead
to asymptotic self-oscillatory states. This behavior is reminiscent
of superfluid time crystals \cite{Autti18,Autti22,Autti21}, where
the conservation of the particle number, together with a non-zero
chemical potential induce robust temporal oscillations in the macroscopic
wave function of the condensed fraction \cite{Boris18,BorisBook}.

Our results indicate that the combination of quantum-optical processes
and condensed-matter models allows one to go beyond the paradigm accessible
to these disciplines separately.

The paper is organized as follows. In the next three sections we introduce
the model, the equations, and the main concepts (Bose levels and Bose
surface) that we use to unravel the behavior of the system. In Section
\ref{Section_NonlinearDissipation} we present the results found in
the absence of local linear loss, whose effect we analyze in Section
\ref{Section_LinearDissipation}. Finally, in Section \ref{Section_Implementation}
we explain how to implement our model with state-of-the-art superconducting-circuit
devices.

\section{Model}

We consider $N$ bosonic modes arranged in the nodes of a $d$-dimensional
cubic array\textcolor{black}{, that is, each node has $2d$ neighbors
(so-called }\textit{\textcolor{black}{coordination number}}\textcolor{black}{).
T}he mode indices are parametrized as $\boldsymbol{j}=(j_{1},...,j_{d})$,
with $j_{n}\in\{1,2,...,L\}$ and $N=L^{d}$. \textcolor{black}{Annihilation
operators $\hat{a}_{\boldsymbol{j}}$ are defined at each node, satisfying}
canonical commutation relations, $[\hat{a}_{\boldsymbol{j}},\hat{a}_{\boldsymbol{l}}]=0$
and $[\hat{a}_{\boldsymbol{j}},\hat{a}_{\boldsymbol{l}}^{\dagger}]=\delta_{\boldsymbol{jl}}$.
Apart from the standard chemical potential and nearest-neighbor hopping
terms of the Bose-Hubbard model, we consider pair-driving in the Hamiltonian
that breaks particle-number conservation, but keeping a $Z_{2}$ symmetry
$\hat{a}_{\boldsymbol{j}}\rightarrow-\hat{a}_{\boldsymbol{j}}\,\forall\boldsymbol{j}$,
see Eq. (\ref{hamiltonian}). Such driving must be necessarily accompanied
by pair dissipation. Here we study the case in which the environment
used for driving is common to all modes, such that a single collective
jump operator $\sum_{\boldsymbol{j}}\hat{a}_{\boldsymbol{j}}^{2}$
effects the decay process. In contrast to local decay, we will show
that such type of decay allows for a much richer, flexible, and controllable
spatiotemporal phenomena. In addition, we consider the possibility
of having local decay, typically unavoidable in real setups, as well
as incoherent pumping to compensate it. We will discuss later plausible
implementations of this model in the context of superconducting-circuit
arrays. The master equation describing the evolution of the state
of the system $\hat{\rho}$ reads
\begin{align}
\partial_{t}\hat{\rho}=-\mathrm{i}\biggl[\frac{\hat{H}}{\hbar},\hat{\rho}\biggr] & +\frac{\gamma}{2N}\mathcal{D}_{\sum_{\boldsymbol{j}}\hat{a}_{\boldsymbol{j}}^{2}}\left[\hat{\rho}\right]\label{MasterEq}\\
 & +\sum_{\boldsymbol{j}}\left(\gamma_{0}\mathcal{D}_{\hat{a}_{\boldsymbol{j}}}\left[\hat{\rho}\right]+\Gamma\mathcal{D}_{\hat{a}_{\boldsymbol{j}}^{\dagger}}\left[\hat{\rho}\right]\right),\nonumber 
\end{align}
where
\begin{equation}
\frac{\hat{H}}{\hbar}=\sum_{\boldsymbol{j}}\left[-\mu\hat{a}_{\boldsymbol{j}}^{\dagger}\hat{a}_{\boldsymbol{j}}+\frac{\varepsilon}{2}\left(\hat{a}_{\boldsymbol{j}}^{2}+\hat{a}_{\boldsymbol{j}}^{\dagger}{}^{2}\right)\right]-\mathinner{\color{black}\frac{J}{2d}}\sum_{\langle\boldsymbol{jl}\rangle}\hat{a}_{\boldsymbol{j}}\hat{a}_{\boldsymbol{l}}^{\dagger},\label{hamiltonian}
\end{equation}
and
\begin{equation}
\mathcal{D}_{A}[\hat{\rho}]=2\hat{A}\hat{\rho}\hat{A}^{\dagger}-\hat{A}^{\dagger}\hat{A}\hat{\rho}-\hat{\rho}\hat{A}^{\dagger}\hat{A}.\label{LindbladForm}
\end{equation}
Here $\mu$ plays the role of a chemical potential (corresponding
in the implementation to the detuning of a driving field, which is
fully tunable, as we will see later), $\varepsilon$ is the pair-injection
rate, and $J$ is the hopping rate \textcolor{black}{(normalized to
the coordination number for convenience)}, the $\langle\boldsymbol{jl}\rangle$
symbol denoting that the sum runs only over nearest neighbors. As
for the parameters in the incoherent terms, all assumed of the Lindblad
form (\ref{LindbladForm}) under the standard Born-Markov conditions
satisfied by most quantum-optical systems, $\gamma_{0}$ is the local
linear decay rate, $\Gamma$ is the pumping rate, and $\gamma/2N$
is the collective nonlinear decay rate, which is divided by $N$ in
order to make all terms in the master equation extensive. All the
parameters are taken positive except for $\mu$, which we allow to
take on negative values, as this is naturally the case in experiments
(the driving can be red-detuned or blue-detuned with respect to resonance).
\textcolor{black}{Note that (\ref{MasterEq}) is quartic in the bosonic
operators, through $\mathcal{D}_{\sum_{\boldsymbol{j}}\hat{a}_{\boldsymbol{j}}^{2}}$,
and hence it is a many-body master equation with no Gaussian or simple
analytic solution.}

We assume to be deep in the superfluid \textcolor{black}{phase \cite{BorisBook}},
where the $Z_{2}$ symmetry is spontaneously broken and the system
is well described by a coherent-state ansatz with amplitudes $\langle\hat{a}_{\boldsymbol{j}}\rangle=\psi_{\boldsymbol{j}}\in\mathbb{C}$.
Let us remark that, while one might argue that driving and dissipation
would prevent the system to reach true superfluidity, it is by now
well established that \textcolor{black}{superfluid order exists even
in driven-dissipative systems for sufficiently large $J$}. In particular,
both \textcolor{black}{theory/simulations \cite{Kinsler91,Greentree06,Noh17,Navarrete17cycles,Iemini18}}
and experiments \cite{Carusotto13rev,Giacobino20rev,JaqBloch20rev}
show that it takes an infinite time to tunnel in between the symmetry-breaking
coherent states in the thermodynamic limit for sufficiently weak interactions
or nonlinearity, and hence such superfluid states are true robust
asymptotic states accessible to \textcolor{black}{driven-dissipative
systems}. As shown in Appendix \ref{App:MEtoGP}, in this regime the
master equation (\ref{MasterEq}) is then translated into a set of
nonlinear differential equations for the coherent amplitudes (generalized
Gross-Pitaevskii or GP equations). Furthermore, we can set two parameters
to one, say $J=1$ and $\gamma/N=1$, which is equivalent to normalizing
all rates to $J$, time to $J^{-1}$, and the amplitudes $\psi_{\boldsymbol{j}}$
to $\sqrt{NJ/\gamma}$, as detailed in Appendix \ref{App:MEtoGP}.
The resulting equations read
\begin{equation}
\dot{\psi}_{\boldsymbol{j}}=\left(\mathrm{i}\mu-\kappa\right)\psi_{\boldsymbol{j}}-\left(\mathrm{i}\varepsilon+\sum_{\boldsymbol{l}}\psi_{\boldsymbol{l}}^{2}\right)\psi_{\boldsymbol{j}}^{*}+\mathinner{\color{black}\frac{\mathrm{i}}{2d}}\sum_{\left<\boldsymbol{l}\right>_{\boldsymbol{j}}}\psi_{\boldsymbol{l}},\label{GPrealspace}
\end{equation}
where $\left<\boldsymbol{l}\right>_{\boldsymbol{j}}$ denotes
that the sum extends only over nearest neighbors of $\boldsymbol{j}$,
and we have defined the \textit{effective linear decay rate} $\kappa=\gamma_{0}-\Gamma$,
which combines the local damping and pumping, and is assumed to be
positive or zero in the following (that is to say, pumping can only
compensate damping, but not go above it).

Owed to the translational invariance of the problem (we assume periodic
boundary conditions for simplicity), it is more convenient to work
in the Fourier basis 
\begin{equation}
\alpha_{\boldsymbol{k}}=\frac{1}{\sqrt{N}}\sum_{\boldsymbol{j}}e^{\mathrm{i}\boldsymbol{k}\cdot\boldsymbol{j}}\psi_{\boldsymbol{j}},
\end{equation}
with $\boldsymbol{k}=(k_{1},...,k_{d})$ the Fourier wave vector,
whose components take on values
\begin{equation}
\frac{k_{n}L}{2\pi}\in\left\{ -\left\lfloor \frac{L}{2}\right\rfloor ,-\left\lfloor \frac{L}{2}\right\rfloor +1,...,-\left\lfloor \frac{L}{2}\right\rfloor +L-1\right\} ,
\end{equation}
working in the first Brillouin zone, meaning that the Fourier indices
become continuous $k_{n}\in[-\pi,\pi[$ in the thermodynamic limit
$L\rightarrow\infty$. Here the floor operation $\bigl\lfloor x\bigr\rfloor$
provides the greatest integer less than or equal to $x$. Transforming
the GP equations (\ref{GPrealspace}) to Fourier space, we obtain
\begin{equation}
\dot{\alpha}_{\boldsymbol{k}}=\left(\mathrm{i}\omega_{\boldsymbol{k}}-\kappa\right)\alpha_{\boldsymbol{k}}-\left(\mathrm{\mathrm{i}}\varepsilon+\sum_{\boldsymbol{q}}\alpha_{\boldsymbol{q}}\alpha_{-\boldsymbol{q}}\right)\alpha_{-\boldsymbol{k}}^{*},\label{GPequation}
\end{equation}
where
\begin{equation}
\omega_{\boldsymbol{k}}=\mu+\mathinner{\color{black}\frac{1}{d}}\sum_{n=1}^{d}\cos k_{n},
\end{equation}
is the dispersion relation (with inverted sign for convenience),
which we represent in Fig. \ref{Fig-Dispersions} for dimensions $d=1$
and $2$.

Interestingly, these GP equations are invariant under several transformations,
most notably: (i) continuous shifts of the relative phase between
each $\pm\boldsymbol{k}$ pair, (ii) continuous time translations,
and (iii) the exchange $\pm\boldsymbol{k}_{1}\rightleftharpoons\pm\boldsymbol{k}_{2}$
of pairs with the same value of the dispersion relation, $\omega_{\boldsymbol{k}_{1}}=\omega_{\boldsymbol{k}_{2}}$,
which is also a continuous symmetry in the thermodynamic limit for
$d>1$. This implies that any symmetry-breaking solution will be accompanied
by the corresponding Goldstone modes whose fluctuations can make the
solutions change without opposition, as we will see later.

This is the set of equations that we focus on in this work. Let us
remark that, while we have performed extensive numerical simulations
of the equations, most of the results are understood from analytical
or semi-analytical arguments, as we explain below. Before proceeding,
though, let us introduce a few concepts that will turn out to be very
useful for the analysis of the asymptotic states.

\section{Bose levels \\and collective equations\label{Section_BoseLevels}}

\textcolor{black}{For the sake of introducing the so-called }\textit{\textcolor{black}{Bose
levels}}\textcolor{black}{, which will be central to our analysis,
let us first focus on the dynamics induced by the Hamiltonian only.
Removing momentarily the incoherent processes from the GP equations
(\ref{GPequation}), these turn into $\dot{\alpha}_{\boldsymbol{k}}=\mathrm{i}\omega_{\boldsymbol{k}}\alpha_{\boldsymbol{k}}-\mathrm{i}\varepsilon\alpha_{-\boldsymbol{k}}^{*}$.}
When $\varepsilon<|\omega_{\boldsymbol{k}}|$, the amplitudes $\alpha_{\boldsymbol{k}}(t)$
remain bounded and oscillate in time at frequency $\sqrt{\omega_{\boldsymbol{k}}^{2}-\varepsilon^{2}}$.
In contrast, when $\varepsilon>|\omega_{\boldsymbol{k}}|$ these frequencies
become imaginary, meaning that the amplitudes diverge exponentially
with time at rate $\sqrt{\varepsilon^{2}-\omega_{\boldsymbol{k}}^{2}}$.
When including dissipation, we expect that the modes with largest
divergence rate will dominate the long-term dynamics of the model.
When the chemical potential satisfies \textcolor{black}{$|\mu|<1$}
(which we assume in the following), the most divergent modes are characterized
by $\omega_{\boldsymbol{k}}=0$, and define the bosonic analog of
the Fermi surface, which we dubbed the \textit{Bose surface} in \cite{WangCaiNB20}
(highlighted thick black in Fig. \ref{Fig-Dispersions}). In general,
since Fourier modes with the same value of $\omega_{\boldsymbol{k}}$
have the same divergence rate or oscillation frequency, it is convenient
to group them into what we will call \textit{Bose levels} (see Fig.
\ref{Fig-Dispersions}). Let us denote by $\{\beta\}$ the collection
of distinct values that $\omega_{\boldsymbol{k}}$ can take, which
take continuously over the whole interval \textcolor{black}{$[\mu-1,\mu+1]$}
in the thermodynamic limit, but otherwise form a discrete set within
that interval. We define Bose level $\beta$, and abbreviate it by
$\mathrm{BL}_{\beta}$, as the set of Fourier modes $\boldsymbol{k}$
for which $\omega_{\boldsymbol{k}}=\beta$. The particular Bose level
with $\beta=0$, $\mathrm{BL}_{0}$, is then the Bose surface (abbreviated
$\mathrm{BS}$), containing the most divergent modes. Note that the
Bose levels with $\beta={\color{black}\mu\pm1}$ lying at the center
and the edge of the Brillouin zone contain a single mode, respectively,
$\boldsymbol{k}=(0,...,0)$ or $(\pi,...,\pi)$; none of these two
Bose levels are the Bose surface, because of our previous ${\color{black}|\mu|<1}$
assumption. All other Bose levels form a closed curve ($d=2$) or
surface ($d=3$) in Fourier space, and are constituted by multiple
$\pm\boldsymbol{k}$ pairs with opposite wave vector, see Fig. \ref{Fig-Dispersions}(b).
This does not hold in one dimension ($d=1$), for which Bose levels
are formed by a single pair of opposite wave-vector modes, see Fig.
\ref{Fig-Dispersions}(a).

\begin{figure*}[t]
\includegraphics[width=1\textwidth]{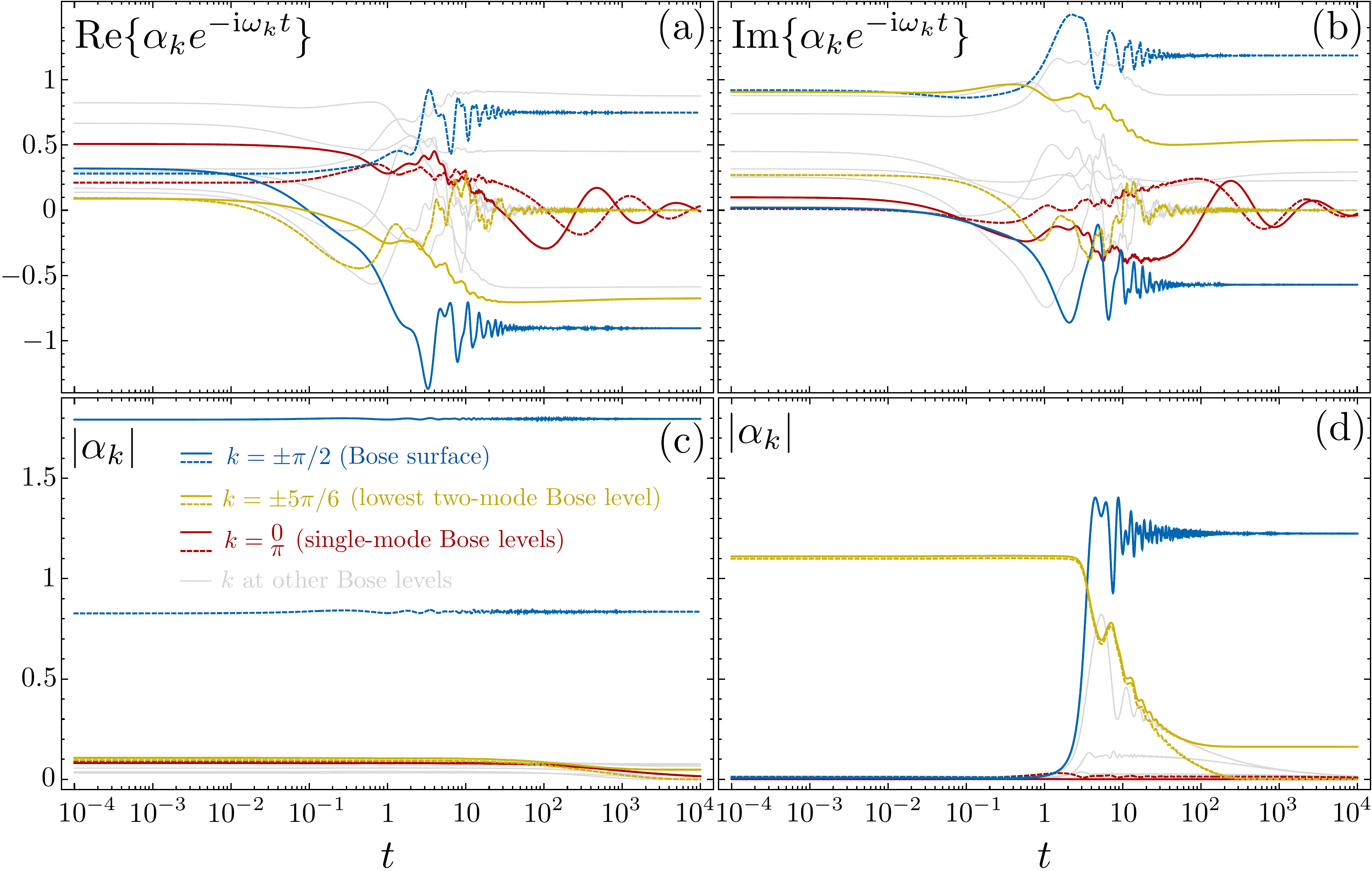} \caption{Numerical simulation of the time evolution of $\alpha_{k}(t)$ from
the Gross-Pitaevskii equations (\ref{GPequation}) in 1D ($d=1$)
for size $L=12$. Here we consider no local effective damping ($\kappa=0$),
taking $\varepsilon=3$ and $\mu=0$, but the same behavior is found
for other parameters, sizes, and dimensions. We highlight the modes
$k=\pm\pi/2$ at the Bose surface $\omega_{\pi/2}=0$ (blue), the
modes $k=0$ and $\pi$ at the edge of the dispersion relation $\omega_{0,\pi}={\color{black}\mu\pm1}$
(red), and the modes $k=\pm5\pi/6$ at the lowest nontrivial Bose
level (yellow). The thinner grey lines correspond to the rest of the
modes. (a) and (b) show the real and imaginary parts of $\alpha_{k}e^{-\mathrm{i}\omega_{k}t}$
starting from random initial conditions. Note that the lines converge
to steady values, showing that the $k$ modes oscillate as $e^{\mathrm{i}\omega_{k}t}$
in the asymptotic state. Most population flows into the modes at the
Bose surface, but one mode of each $\pm k$ pair at other Bose levels
remains populated, as required by the existence of the constants of
motion that we discuss in the main text. The population of modes $k=0$
and $\pi$ vanishes asymptotically, but note that their relaxation
rate can be much slower than that of the other Bose levels. (c) shows
the populations $|\alpha_{k}|$ starting from a random perturbation
($\protect\leq$5\%) around the stationary solution (\ref{BSstatSol})
with only the Bose-surface modes excited, specifically the imbalanced
configuration $\alpha_{\pi/2}=\sqrt{\varepsilon}e^{\mathrm{i}\pi/4}=2\alpha_{-\pi/2}$.
The simulations agree with these stationary solutions being (marginally)
stable. In contrast, (d) starts from a random perturbation ($\protect\leq$1\%)
around a stationary solution (\ref{BLstatSol}) with only the $k=\pm5\pi/6$
modes excited, and shows that such stationary solutions are very unstable.
All these numerical simulations (and the many more we have conducted,
including 2D and 3D situations) agree with the analytical results
that we work out in the main text.}
\label{Fig-ConfigEvo}
\end{figure*}

\textcolor{black}{With the definition of Bose levels at hand, we can
now go back to the complete GP equations (\ref{GPequation}), including
incoherent terms. These equations are simplified by defining the}
\textit{collective Bose-level variables}
\begin{subequations}\label{CollectiveVariables}
\begin{align}
\sum_{\boldsymbol{k}\in\mathrm{BL}_{\beta}}\alpha_{\boldsymbol{k}}\alpha_{\boldsymbol{-k}} & \equiv s_{\beta},\\
\sum_{\boldsymbol{k}\in\mathrm{BL}_{\beta}}\left|\alpha_{\boldsymbol{k}}\right|^{2} & \equiv n_{\beta},
\end{align}
\end{subequations}
which wash off the details about each specific
mode of $\mathrm{BL}_{\beta}$, but describe the excitation of the
level as a whole. In particular, note that $n_{\beta}$ provides the
total population at $\mathrm{BL}_{\beta}$. It is simple, using the
GP equations (\ref{GPequation}), to show that the collection of these
Bose-level variables evolve according to a closed set of equations
that reads
\begin{subequations}\label{BoseLevelEOM}
\begin{align}
\dot{s}_{\beta} & =2(\mathrm{i}\beta-\kappa)s_{\beta}-2n_{\beta}\Bigl(\mathrm{i}\varepsilon+\sum_{\beta'}s_{\beta'}\Bigr),\label{ds}\\
\dot{n}_{\beta} & =-2\kappa n_{\beta}-\Bigl[s_{\beta}^{*}\Bigl(\mathrm{i}\varepsilon+\sum_{\beta'}s_{\beta'}\Bigr)+\mathrm{c.c.}\Bigr],\label{dz}
\end{align}
\end{subequations}
where the sums extend over all Bose levels. We
dub these the Bose-level equations of motion. They will be very useful
when studying the types of asymptotic states that the system can reach,
which we pass now to discuss first for $\kappa=0$ in the next section,
and then considering the effect of linear effective decay $\kappa\neq0$
one section later.

\newpage
\section{Purely nonlinear dissipation: \\constants of motion \\and oscillatory states\label{Section_NonlinearDissipation}}

In this section we analyze the long-term solutions of the GP equations
in the case of a perfect balance between linear gain and loss, $\Gamma=\gamma_{0}$,
that is, absence of linear effective decay, $\kappa=0$. We concentrate
first on stationary solutions, setting $\dot{\alpha}_{\boldsymbol{k}}=0$
in (\ref{GPequation}). Note that the trivial state $\alpha_{\boldsymbol{k}}=0$
exists but it's unstable by assumption: since we assume that the $\omega_{\boldsymbol{k}}=0$
Bose surface exists, even an infinitesimal driving rate $\varepsilon$
will make fluctuations increase exponentially towards a nontrivial
solution (see Appendix \ref{App:TrivialStabilityAnalysis}). On the
other hand, we prove in Appendix \ref{App:StationarySols} that stationary
solutions can populate only modes with the same $|\omega_{\boldsymbol{k}}|$.
Hence, stationary solutions only populate at most two Bose levels,
those at opposite values of the dispersion relation, say, $\omega_{\boldsymbol{k}}=\pm\Omega$.

When the populated level is the Bose surface, that is, when all $\alpha_{\boldsymbol{k}\notin\mathrm{BS}}=0$,
we can set $\omega_{\boldsymbol{k}}=0$ in the GP equations (\ref{GPequation})
and obtain a set of stationary solutions constrained only by the condition
\begin{equation}
\sum_{\boldsymbol{k}\in\mathrm{BS}}\alpha_{\boldsymbol{k}}\alpha_{\boldsymbol{-k}}=-\mathrm{i}\varepsilon.\label{BSstatSol}
\end{equation}
Any configuration of amplitudes $\alpha_{\boldsymbol{k}\in\mathrm{BS}}$
satisfying this condition is a valid stationary solution. We comment
on the stability of these configurations shortly, but let us anticipate
that they are marginally stable, meaning that fluctuations around
them do not grow, but might not be damped either, see Fig. \ref{Fig-ConfigEvo}(c).
This is owed to Goldstone's theorem applied to the continuous symmetries
of the GP equations that are spontaneously broken by the stationary
solution, as well as to the existence of extra constants of motion
that we introduce below.

Let us consider now stationary configurations populating an arbitrary
pair of Bose levels with $\omega_{\boldsymbol{k}}=\pm\Omega$, which
we denote here by $\mathrm{BL}_{\pm}\neq\mathrm{BS}$, so that $\alpha_{\boldsymbol{k}\notin\mathrm{BL_{\pm}}}=0$.
In contrast to the Bose-surface solutions, we will prove later and
show in Fig. \ref{Fig-ConfigEvo}(d) that these ones are unstable.
Nevertheless, it is still instructive to consider and understand them,
as they will allow us to introduce some useful concepts, as well as
provide insight into the privileged role of the Bose surface. In Appendix
\ref{App:StationarySols} we discuss this type of stationary configurations
in detail. They are constrained by two conditions
\begin{subequations}\label{BLstatSol}
\begin{align}
|n_{+}-n_{-}| & =\sqrt{\varepsilon^{2}-\Omega^{2}},\\
\alpha_{\boldsymbol{-k}\in\mathrm{BL}_{\pm}} & =-\mathrm{i}e^{\mathrm{i}\varphi_{\pm}}\alpha_{\boldsymbol{k}}^{*},\label{BLstatSol(b)}
\end{align}
\end{subequations}
where $\varphi_{\pm}=\arg\left\{ \pm(n_{+}-n_{-}+\mathrm{i}\Omega)\right\} $
is a phase common to all $\boldsymbol{k}$ modes within the same Bose
level. The first thing to note is that only Bose levels for which
$\varepsilon>|\Omega|$ can be excited as stationary solutions. Remarkably,
only the difference in populations $|n_{+}-n_{-}|$ is fixed, but
the total population $n_{+}+n_{-}$ is arbitrary, except for the algebraic
constraint $n_{+}+n_{-}\geq|n_{+}-n_{-}|$. Moreover, expressing the
amplitudes in magnitude and phase as $\alpha_{\boldsymbol{k}}=\rho_{\boldsymbol{k}}e^{\mathrm{i}\phi_{\boldsymbol{k}}}$,
the second condition (\ref{BLstatSol(b)}) means that modes with opposite
wave vector must be equally populated, $\rho_{\boldsymbol{k}}=\rho_{-\boldsymbol{k}}$,
with their phase sum fixed to a value common to all $\pm\boldsymbol{k}$
pairs of the same level, $\phi_{\boldsymbol{k}}+\phi_{-\boldsymbol{k}}=\varphi_{\beta}-\pi/2$
in this case; we will denote this by \textit{balanced} solutions.

Before discussing the stability of these solutions, we need to talk
about constants of motion. Setting $\kappa=0$ in (\ref{BoseLevelEOM}),
it is easy to see that the combination 
\begin{equation}
C_{\beta}=\sqrt{n_{\beta}^{2}-|s_{\beta}|^{2}},\label{Cbeta}
\end{equation}
of Bose-level variables are constants of motion, $\dot{C}_{\beta}=0$;
we dub them \textit{level constants}. In order to understand the physical
meaning of these constants, let us first consider the $d=1$ case
(one-dimensional system), for which there are only two modes $\pm k$
at a generic $\mathrm{BL}_{\beta}$, so that it is straightforward
to see that $C_{\beta}=\bigl||\alpha_{k}|^{2}-|\alpha_{-k}|^{2}\bigr|$.
Hence, these level constants are a measure of the imbalance between
opposite-momenta modes present in the solution. As we show in Appendix
\ref{App:LevelConstants}, the same conclusion holds in higher dimensions
$d>1$: $C_{\beta}=0$ if and only if the configuration of the amplitudes
$\alpha_{\boldsymbol{k}}$ at that Bose level $\mathrm{BL}_{\beta}$
is balanced in the sense explained in the previous paragraph. The
existence of these level constants is crucial in order to understand
the stability analysis of the aforementioned stationary solutions,
as well as the existence of oscillatory asymptotic states.

\begin{figure*}[t]
\includegraphics[width=1\textwidth]{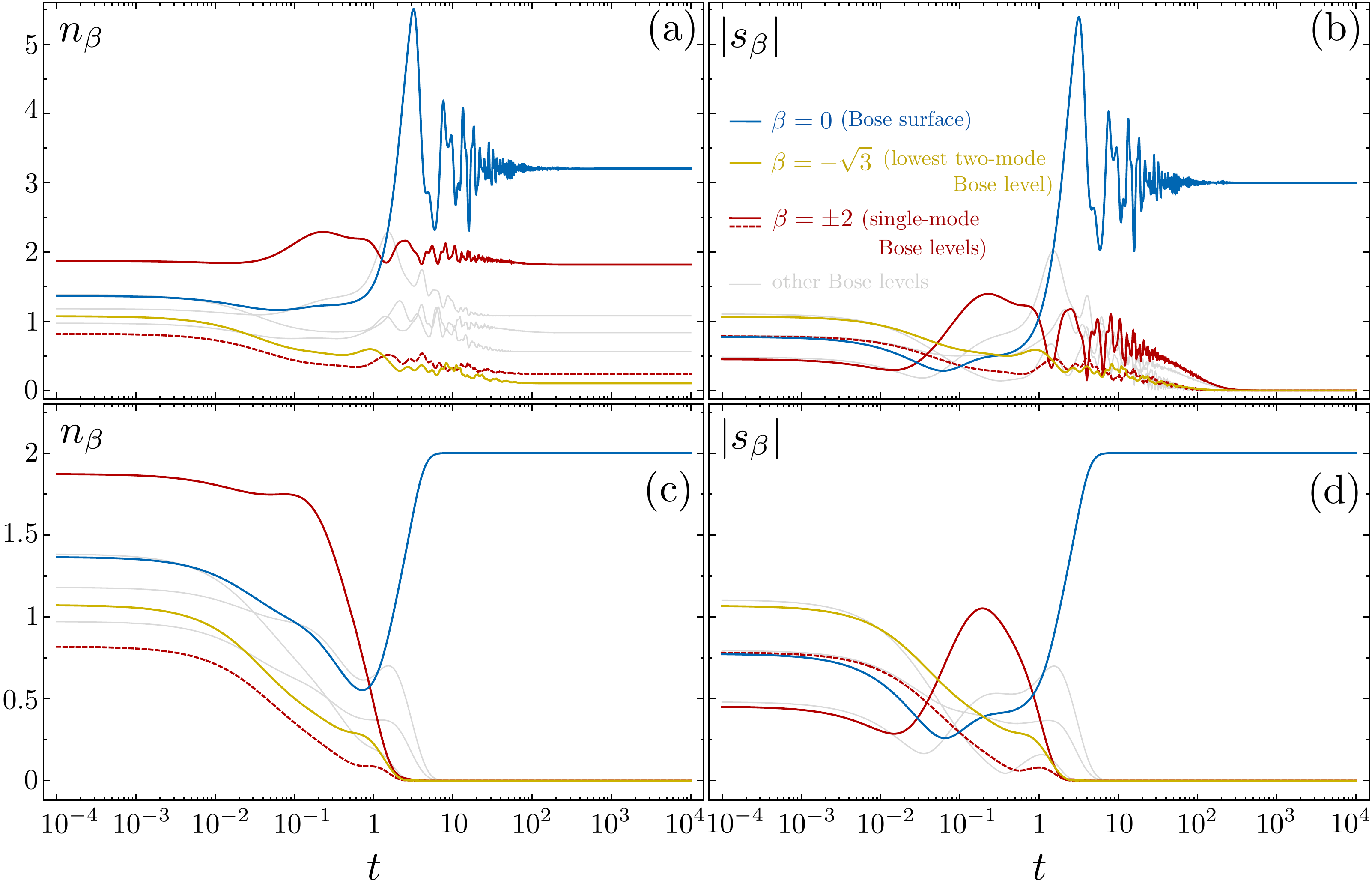} \caption{Numerical time-evolution of the Bose-level variables $n_{\beta}=\sum_{\boldsymbol{k}\in\mathrm{BL}_{\beta}}|\alpha_{\boldsymbol{k}}|^{2}$
and $s_{\beta}=\sum_{\boldsymbol{k}\in\mathrm{BL}_{\beta}}\alpha_{\boldsymbol{k}}\alpha_{-\boldsymbol{k}}$,
which form a closed system ruled by Eqs. (\ref{BoseLevelEOM}). As
in Fig. \ref{Fig-ConfigEvo}, we consider a 1D situation with $L=12$,
$\mu=0$, and $\varepsilon=3$, but the plots are qualitatively the
same for any other choice of dimension, size, or parameters. We consider
in (a) and (b) the $\kappa=0$ case, and in (c) and (d) the $\kappa=1$
case, starting from a random initial state. Again as in Fig. \ref{Fig-ConfigEvo},
we highlight the evolution of the Bose-surface (blue), the lowest
nontrivial Bose level $\mathrm{BL}_{-\sqrt{3}}$ containing modes
$k=\pm5\pi/2$ (yellow), and the Bose levels ${\color{black}\mathrm{BL}_{\pm1}}$
with the largest and smallest values of the dispersion relation, which
contain the single modes $k=0$ (solid red) and $\pi$ (dashed red),
respectively. The simulations are consistent with the analytics discussed
in the main text, showing that all population moves towards the Bose
surface, except when $\kappa=0$, for which a population $\lim_{t\rightarrow\infty}n_{\beta\protect\neq0}(t)=C_{\beta}$
persists, associated with the oscillatory configurations discussed
in Fig. \ref{Fig-ConfigEvo} induced by the existence of the $C_{\beta}=\sqrt{n_{\beta}^{2}-|s_{\beta}|^{2}}$
constants of motion.}
\label{Fig-LevelsEvo}
\end{figure*}

In order to see how the constants of motion imply the existence of
non-stationary stable solutions, we simply remind that stationary
solutions cannot have Bose levels with different $|\beta|$ excited
simultaneously, and moreover, stationary solutions at Bose levels
other than the Bose surface are balanced, meaning that $C_{\beta\neq0}=0$
for such stationary configurations. On the other hand, we have just
seen that the level constants $C_{\beta}$ are conserved and are a
measure for how imbalanced the configuration of amplitudes $\alpha_{\boldsymbol{k}}$
is. Hence, starting with $C_{\beta\neq0}\neq0$, it is not possible
that all the $\alpha_{\boldsymbol{k}}$ of $\mathrm{BL}_{\beta}$
are zero simultaneously, which implies that the asymptotic amplitudes
$\lim_{t\rightarrow\infty}\alpha_{\boldsymbol{k}}(t)$ cannot be stationary
for all $\boldsymbol{k}\in\mathrm{BL}_{\beta}$, and in particular
oscillate in time at frequency $\omega_{\boldsymbol{k}}$ as shown
in Figs. \ref{Fig-ConfigEvo}(a,b) and discussed below. This is similar
to what happens in superfluid time crystals \cite{Autti18,Autti21,Autti22},
for which particle-number conservation forces the superfluid wave
function to oscillate with a frequency equal to the chemical potential
\cite{Boris18,BorisBook}. Interestingly, even if the individual
amplitudes $\alpha_{\boldsymbol{k}}(t)$ can oscillate, our extensive
simulations show that the Bose-level variables, ruled by Eqs. (\ref{BoseLevelEOM}),
reach the fixed point 
\begin{align}
\left\{ \begin{array}{cc}
n_{\beta}=C_{\beta},\,s_{\beta}=0,\qquad & \text{if }\mathrm{BL}_{\beta}\neq\mathrm{BS}\\
n_{\beta}=\sqrt{C_{\beta}^{2}+\varepsilon^{2}},\,s_{\beta}=-\mathrm{i}\varepsilon, & \text{if }\mathrm{BL}_{\beta}=\mathrm{BS}
\end{array}\right.,\label{BLeqsFixedPoint}
\end{align}
from any initial condition, as shown in Figs. \ref{Fig-LevelsEvo}(a,b).
Incidentally, this shows that the population induced by the driving
$\varepsilon$ concentrates at the Bose surface, while the rest of
Bose levels only keep the minimum amount of population required to
satisfy the conservation of the level constant $C_{\beta}$. It is
interesting to analyze the configuration of the amplitudes $\alpha_{\boldsymbol{k}}$
in this oscillatory regime. As proven analytically in Appendix \ref{App:OasciSols},
at a given $\mathrm{BL}_{\beta}\neq\mathrm{BS}$ all modes oscillate
at frequency $\beta$, but, remarkably, at least one amplitude of
each $\pm\boldsymbol{k}$ pair must vanish, see also Fig. \ref{Fig-ConfigEvo}.
The Bose-surface modes have $\beta=0$, so they can remain stationary,
and are only constrained by (\ref{BLeqsFixedPoint}).

Once we understand that there is a part of the population of the Bose
levels (associated with $C_{\beta}\neq0$ ) that the dynamics cannot
get rid of, we can discuss the stability of the stationary solutions.
Let's start with the stationary solution (\ref{BSstatSol}), corresponding
to a stationary configuration of the amplitudes at the Bose surface.
From the previous discussion, it is clear that any perturbation that
increases $C_{\beta}$ at other Bose levels will leave a remanent
of population in those levels that will make the state start oscillating.
We show this analytically in Appendix \ref{App:BSGPstabilityAnalysis}
by studying the stability from the GP equations (\ref{BoseLevelEOM}).
However, it is crucial to understand that if the perturbation is small,
it will remain small (will not grow exponentially), and hence the
oscillations will just be a small perturbation on top of the stationary
solution, as shown in Fig. \ref{Fig-ConfigEvo}(c). In addition, note
that (\ref{BSstatSol}) only fixes the sum $\sum_{\boldsymbol{k}\in\mathrm{BS}}\alpha_{\boldsymbol{k}}\alpha_{-\boldsymbol{k}}$,
which means that the part of the fluctuations that changes the phase
difference between each $\pm\boldsymbol{k}$ pair, the ratio between
their magnitudes, or even the distribution of population among the
modes of the Bose surface, will not damp back to the original configuration,
but will also not move away exponentially. This is an effect derived
from the Goldstone theorem applied to the continuous symmetries present
in the GP equations (\ref{GPequation}), as we anticipated when we
introduced the equations. This implies that the linear stability matrix
associated to this solution is plagued with eigenvalues with zero
real part, but none with positive real part. In this sense, this stationary
solution is only marginally stable, but still, it is not unstable.

The story is completely different for the stationary solutions at
other Bose levels. As we prove in Appendix \ref{App:BoseLevelStabilityAnalysis}
and show in Fig. \ref{Fig-ConfigEvo}(d), for the stationary solution
(\ref{BLstatSol}) at a given Bose-level even an infinitesimal fluctuation
at Bose levels with smaller $|\beta|$ will precipitate a cascade
of the population towards the lowest-$|\beta|$ level, that is, towards
the Bose surface. The cascade proceeds until just the small population
$n_{\beta}=\delta C_{\beta}$ related to the level constant induced
by the initial fluctuations, $\delta C_{\beta}^{2}=n_{\beta}^{2}(0)-|s_{\beta}(0)|^{2}$,
remains in Bose levels other than the Bose surface. Hence, these stationary
solutions are very unstable, and moreover reinforce the idea that
the Bose-surface solution is (marginally) stable.

In summary, in the absence of linear effective dissipation ($\kappa=0$),
the asymptotic configuration of the system corresponds to one in which
the population is concentrated at the Bose surface in the form of
a stationary solution with $n_{0}=\sqrt{C_{0}^{2}+\varepsilon^{2}}$,
except for some leftover population $n_{\beta\neq0}=C_{\beta}=\sqrt{n_{\beta}^{2}(0)-|s_{\beta}(0)|^{2}}$
at other Bose levels, for which the amplitudes $\alpha_{\boldsymbol{k}\in\mathrm{BL}_{\beta}}$
oscillate in time at frequency $\beta$. The specific way in which
the population is distributed over the $\boldsymbol{k}$ modes of
a given Bose level is only constrained by $s_{0}=-\mathrm{i}\varepsilon$
at the Bose surface and $\alpha_{\boldsymbol{k}}\alpha_{-\boldsymbol{k}}=0$
at other Bose levels.

\section{Effect of linear effective dissipation\protect\label{Section_LinearDissipation}}

With the previous results in mind, it is easy to understand what happens
when introducing linear effective dissipation $\kappa\neq0$, that
is, an imbalance between linear gain $\Gamma$ and loss $\gamma_{0}$.
As we argue next, the main effect of $\kappa$ is making the stationary
configurations at the Bose surface even more robust, that is, to remove
any trace of population in other Bose levels. But, remarkably, even
under these conditions the way in which the population is distributed
along the Bose surface remains arbitrary, and it can be biased with
a judicious initial condition or with a weak drive, as we show below.
In particular, this allows for configurations in which all the $\boldsymbol{k}$
modes of the closed manifold are populated. This is an exotic state
of outmost importance in condensed-matter physics for its connection
with high-$T_{\text{c}}$ superconductivity \cite{Jiang2019}, frustrated
magnetism \cite{Sedrakyan2015}, and interacting problems with spin-orbit
coupling \cite{Wu2011,Gopalakrishnan2011}.

Let us start by noting that, using Eqs. (\ref{BoseLevelEOM}), the
level constants are easily shown to satisfy
\begin{equation}
\dot{C}_{\beta}=-2\kappa C_{\beta}\Longrightarrow\lim_{t\rightarrow\infty}C_{\beta}(t)=0\,\forall\beta.\label{ZeroC}
\end{equation}
The level constants are not conserved anymore; instead, they are forced
to vanish asymptotically by the local dissipation. This is accompanied
by two effects. First, now that the level constants are zero, there
is no reason for any population to remain at Bose levels other than
the Bose surface. Indeed, our exhaustive simulations show that all
population flows towards the Bose surface asymptotically, irrespective
of the initial state, as shown in Figs. \ref{Fig-LevelsEvo}(c,d).
In particular, the system reaches a fixed point of the Bose-level
equations (\ref{BoseLevelEOM}) with
\begin{equation}
\left\{ \begin{array}{cc}
n_{\beta}=s_{\beta}=0,\qquad & \text{if }\mathrm{BL}_{\beta}\neq\mathrm{BS}\\
n_{\beta}=\varepsilon-\kappa=\mathrm{i}s_{\beta}, & \text{if }\mathrm{BL}_{\beta}=\mathrm{BS}
\end{array}\right..
\end{equation}
Note that now it is necessary that $\varepsilon>\kappa$ for this
solution to exist, also required in order to destabilize the trivial
solution $\alpha_{\boldsymbol{k}}=0\,\forall\boldsymbol{k}$, whose
fluctuations will only grow once driving provides enough energy to
balance dissipation, as explicitly shown in Appendix \ref{App:TrivialStabilityAnalysis}.
The flow of the population towards the Bose surface is further supported
by the stability analysis of the stationary solutions at Bose levels
different than the Bose surface, which we detail in Appendices \ref{App:StationarySols}
and \ref{App:BoseLevelStabilityAnalysis}, and proves that these solutions
become more unstable the larger $\kappa$ is. In contrast, the stationary
solution at the Bose surface becomes more stable by increasing $\kappa$,
as we show next.

\begin{figure}[b]
\includegraphics[width=1\columnwidth]{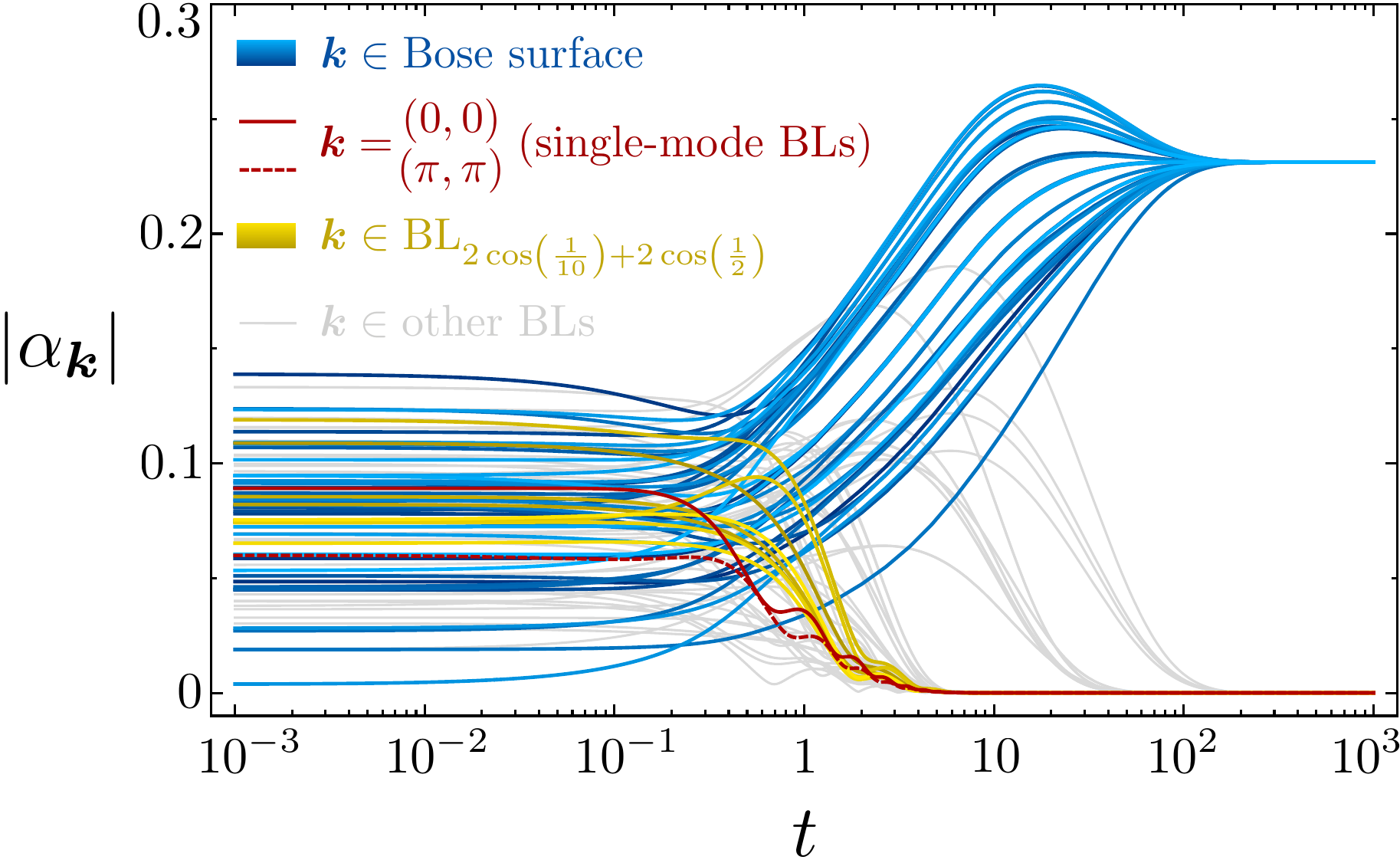} \caption{Numerical simulation of the time evolution of $\alpha_{\boldsymbol{k}}(t)$
from the Gross-Pitaevskii equations (\ref{GPequation}) in 2D ($d=2$)
for a lattice of size $L\times L=20\times20$. We have included a
weak driving term $\sum_{\boldsymbol{k}\in\mathrm{BS}}\mathrm{i}F(\hat{b}_{\boldsymbol{k}}^{\dagger}-\hat{b}_{\boldsymbol{k}})$
with $F=0.01$, where $\hat{b}_{\boldsymbol{k}}=\sum_{\boldsymbol{j}}e^{\mathrm{i}\boldsymbol{k}\cdot\boldsymbol{j}}\hat{a}_{\boldsymbol{j}}/L$
are the annihilation operators for the Fourier modes. The rest of
parameters are $\kappa=1$, $\mu=0$, and $\varepsilon=3$, leading
to a Bose surface with 38 modes (blue). We also show the single-mode
Bose levels at the edge and center of the Brillouin zone (red), the
8 modes of the Bose level with $\beta=2\cos(1/10)+2\cos(1/2)$ (yellow),
and the modes of 10 other random Bose surfaces (grey). Starting from
random initial conditions, we find that a weak driving is enough to
bias a final state with all modes of the Bose surface equally populated,
while all other Bose levels remain unpopulated. Moreover, we have
found that driving differently each $\pm\boldsymbol{k}$ pair, one
can distribute the total density in any desired way within the Bose
surface modes. Hence, these simulations show that a weak driving can
be used to ``write'' any unconventional superfluid state with all
population distributed at will along a closed manifold in Fourier
space.}
\label{Fig-Driving}
\end{figure}

\begin{figure*}[t]
\includegraphics[width=0.7\textwidth]{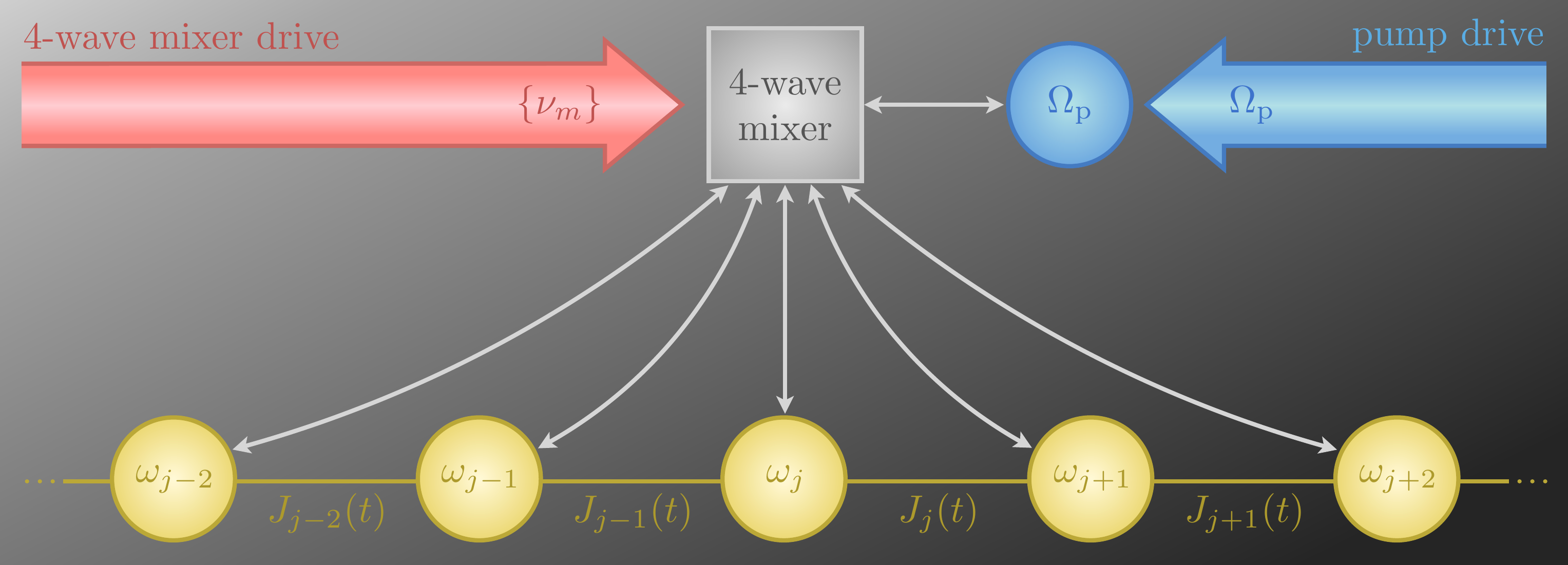} \caption{Schematics of our implementation proposal, especially suited for superconducting
circuits, as detailed in Appendix \ref{App:ATSimplementation}. The main system consists of an array of oscillators (yellow)
with distinct frequencies $\{\omega_{j}\}$, which exchange excitations
at rates that can be tuned at real time, represented by $J_{j}(t)$
in the figure. The array couples through a four-wave mixer to a coherent
polychromatic microwave drive (red) with frequencies $\{\nu_{m}\}$
and an additional lossy oscillator (blue) with frequency $\Omega_{\text{p}}$.
This so-called \textit{pump resonator} is subject to a resonant drive
which, in the absence of coupling to other systems, would induce a
coherent state of amplitude $\alpha_{\mathrm{p}}e^{-\mathrm{i}\Omega_{\mathrm{p}}t}$.
When the decay rate of the pump oscillator is the dominant rate, the
array evolves according to our model's master equation, provided that
the microwave tones $\nu_{m}$, circuit frequencies, and couplings
are chosen appropriately (see the main text).}
\label{Fig-Implementation}
\end{figure*}

The second effect connected to Eq. (\ref{ZeroC}) that we want to
discuss relates to the solution at the Bose surface. As we discussed
in the previous section, a zero level constant implies that the configuration
of the amplitudes $\alpha_{\boldsymbol{k}}=\rho_{\boldsymbol{k}}e^{\mathrm{i}\phi_{\boldsymbol{k}}}$
needs to be of the balanced type. Indeed, solving for the steady state
of the GP equations (\ref{GPequation}) with $\alpha_{\boldsymbol{k}\notin\mathrm{BS}}=0$,
one easily finds for $\boldsymbol{k}\in\mathrm{BS}$ that
\begin{subequations}\label{BSsolLinearDis}
\begin{align}
\phi_{\boldsymbol{k}} & +\phi_{-\boldsymbol{k}}=-\pi/2,\\
\rho_{\boldsymbol{k}} & =\rho_{-\boldsymbol{k}},\quad\sum_{\boldsymbol{k}\in\mathrm{BS}}\rho_{\boldsymbol{k}}^{2}=\varepsilon-\kappa.
\end{align}
\end{subequations}
Any configuration satisfying these conditions is
a robust one that can be biased by, e.g., tuning the initial condition
or adding a weak linear driving with the desired spatial profile (see
Fig. \ref{Fig-Driving}). We say that these configurations are robust
because we show in Appendix \ref{App:BSGPstabilityAnalysis} that
they are stable against any type of perturbation, except of course
perturbations compatible with (\ref{BSsolLinearDis}), which are not
damped but also not amplified (making this solution marginally stable
as expected). Hence, in this system we can write many interesting
patterns, including periodic and quasi-periodic ones with an arbitrary
number of opposite-Fourier modes $\pm\boldsymbol{k}$, as well as
the exotic one presented in Fig. \ref{Fig-Driving} where all the
modes of the Bose surface are populated.

\section{Implementation with superconducting circuit arrays\protect\label{Section_Implementation}}

While our model (\ref{MasterEq}) is potentially implementable in
different experimental platforms, all the elements it requires have
been experimentally demonstrated in superconducting circuits \cite{Blais21,Krantz19,Gu17rev},
which then provide a very flexible platform where our ideas can be
potentially explored. Arrays with up to $N=100$ quantum oscillators
with tunable frequency, anharmonicity, and tunnelings have been already
built \cite{GoogleQAI26_localization,GoogleQAI25,Kim23}, and the
number keeps growing strong in seek of the promised practical quantum
computer. In addition, the pair driving and dissipation considered
in our model has been implemented in single circuits \textcolor{black}{by
exploiting four-wave mixing at driven Josephson junctions \cite{Leghtas15,Touzard18,Markovic18,Leghtas20,Leghtas23,Leghtas24}},
also motivated by the quantum-computing goal of encoding noise-resilient
qubits in the infinite-dimensional Hilbert space of a harmonic oscillator.
Here we show that these experimental quantum computing breakthroughs
can be combined and adapted to implement our pair-driven-dissipative
many-body model (\ref{H0}), as schematically shown in Fig. \ref{Fig-Implementation}.

We consider an array of linear circuits (harmonic oscillators) with
distinct frequencies $\{\omega_{j}\}_{j=1,2,...,N}$ (for notational
simplicity we consider a one-dimensional array). These oscillators
are linearly coupled through an intermediary circuit \textcolor{black}{whose
frequency is sensitive to an external} \textcolor{black}{drive (typically
a magnetic field)}. This allows for real-time tuning of the hopping
rates \cite{Campbell23,Supremacy19,Yan18,Chen14}, which we set to
$J\cos[(\omega_{j}-\omega_{j-1})t]$ between circuits $j-1$ and $j$.\textcolor{black}{{}
Our proposal for the implementation of pair driving and dissipation
relies on the current experimental ability to generate four-wave mixing
\cite{Leghtas15,Touzard18,Markovic18,Leghtas20,Leghtas23,Leghtas24}.
In Appendix \ref{App:ATSimplementation} we provide the details of
a possible concrete circuit design based on the ones in \cite{Leghtas20,Leghtas23,Leghtas24}.
Here, however, we will simply assume that such four-wave mixing process
is available, so we can use it to couple the oscillators of our array
to two elements: the classical coherent tones at frequencies $\{\nu_{m}\}_{m=1,2,...,N}$
fed by a microwave generator and an additional resonantly-driven and
lossy oscillator, dubbed }\textit{\textcolor{black}{pump resonator}}\textcolor{black}{,
with frequency $\Omega_{\mathrm{p}}$ and decay rate $\kappa_{\text{p}}$.
The frequencies are chosen such that only frequency-conversion
processes of the type $\Omega_{\mathrm{p}}+\nu_{n}\rightleftharpoons\omega_{j}+\omega_{l}$
may be near resonant}, so that the evolution of the system's state
$\hat{\rho}_{0}$ is described by the master equation
\begin{equation}
\partial_{t}\hat{\rho}_{0}=\left[\frac{\hat{H}_{0}(t)}{\mathrm{i}\hbar},\hat{\rho}_{0}\right]+\kappa_{\mathrm{p}}\mathcal{D}_{p}\left[\hat{\rho}_{0}\right],\label{H0}
\end{equation}
with (periodic boundaries are assumed to ease the notation, so $j=0$
is understood to be equivalent to $j=N$)
\begin{align}
\frac{\hat{H}_{0}}{\hbar}= & \sum_{j=1}^{N}\left[\omega_{j}\hat{a}_{j}^{\dagger}\hat{a}_{j}-\left(\frac{J}{2}e^{-\mathrm{i}(\omega_{j}-\omega_{j-1})t}\hat{a}_{j-1}\hat{a}_{j}^{\dagger}+\mathrm{H.c.}\right)\right]\nonumber \\
+\Omega_{\text{p}} & \hat{p}^{\dagger}\hat{p}+\sum_{mjl=1}^{N}g\left[e^{-\mathrm{i}\nu_{m}t}\left(\hat{p}+\alpha_{\mathrm{p}}e^{-\mathrm{i}\Omega_{\mathrm{p}}t}\right)\hat{a}_{j}^{\dagger}\hat{a}_{l}^{\dagger}+\mathrm{H.c.}\right],\label{H0_Implementation}
\end{align}
where we have kept only terms that are not suppressed by the rotating
wave approximation (see Appendix \ref{App:ATSimplementation} for
details) under the assumption that $\{\omega_{j}\}$ and $\Omega_{\text{p}}$
are much larger than $J$ and $g$. In this expression, $\alpha_{\mathrm{p}}$
(assumed real and positive without loss of generalization) is the
amplitude of the coherent state that a resonant driving would induce
on the pump resonator in the absence of other processes, such that
the operator $\hat{p}$ annihilates excitations around such coherent
state (see Appendix \ref{App:ATSimplementation}). $g$ is the four-wave-mixing
rate, controllable through the amplitude of the microwave tones \textcolor{black}{(again,
see Appendix \ref{App:ATSimplementation})}, which we assume the same
for all tones. \textcolor{black}{Note that using a simple junction
as the four-wave mixer would generate, in addition to the desired
terms in the second line of (\ref{H0_Implementation}), unwanted or
}\textit{\textcolor{black}{parasitic}}\textcolor{black}{{} terms such
as self-Kerr and cross-Kerr nonlinearities \cite{Leghtas15}. At
low populations, these terms act only as small perturbations \cite{Leghtas15},
whereas at higher populations they can radically alter the predictions
of the previous sections by inducing new nonlinear instabilities.
While these phenomena could be interesting to explore in future work,
for our present purposes it suffices to note that such parasitic terms
can be avoided altogether by replacing the simple junction with designs
based on the so-called asymmetrically threaded SQUID (ATS), as demonstrated
in Refs. \cite{Leghtas20,Leghtas23,Leghtas24}. In Appendix \ref{App:ATSimplementation}
we discuss in detail an ATS-based implementation.}

In order for (\ref{H0}) to lead to the model we have studied, Eq.
(\ref{MasterEq}), the key idea consists in taking far-off resonant
oscillators with $|\omega_{j}-\omega_{l}|\gg g$ and tones with $\nu_{j}=2\omega_{j}-\Omega_{\mathrm{p}}+2\mu$,
where $\mu$ is a small frequency mismatch (detuning) that will play
the role of the chemical potential in the model. \textcolor{black}{This
choice favors the conversion processes $\Omega_{\mathrm{p}}+\nu_{j}\rightleftharpoons2\omega_{j}$,
while suppressing any other one.} This is rigorously seen as follows.
First, let us move to a picture rotating at frequency $\Omega_{\mathrm{p}}$
for the pump and $\omega_{j}+\mu$ for oscillator $j$, effected by
discounting the unitary evolution 
\begin{equation}
\hat{U}(t)=\exp\biggl[-\mathrm{i}\Omega_{\mathrm{p}}t\hat{p}^{\dagger}\hat{p}-\mathrm{i}\sum_{j=1}^{N}(\omega_{j}+\mu)t\hat{a}_{j}^{\dagger}\hat{a}_{j}\biggr].
\end{equation}
The master equation of the transformed (rotating) state $\hat{\rho}_{\text{R}}=\hat{U}^{\dagger}\hat{\rho}_{0}\hat{U}$
keeps the same form as (\ref{Fig-Implementation}), but with a modified
Hamiltonian
\begin{align}
\frac{\hat{H}_{\text{R}}}{\hbar} & =-\sum_{j=1}^{N}\left[\mu\hat{a}_{j}^{\dagger}\hat{a}_{j}+\frac{J}{2}\left(\hat{a}_{j-1}\hat{a}_{j}^{\dagger}+\mathrm{H.c.}\right)\right]\\
 & +\sum_{mjl=1}^{N}g\left[e^{\mathrm{i}\Omega_{mjl}t}\left(\hat{p}+\alpha_{\mathrm{p}}\right)\hat{a}_{j}^{\dagger}\hat{a}_{l}^{\dagger}+\mathrm{H.c.}\right],\nonumber 
\end{align}
with $\Omega_{mjl}=2\omega_{m}-\omega_{j}-\omega_{l}$. It is clear
that $\Omega_{jjj}=0\,\forall j$, and we assume that the frequencies
are chosen in such a way that $|\Omega_{mjl}|\gg\text{max}\{g,g\alpha_{\mathrm{p}}\}$
for any other value of the indices $(m,j,l)$. Then, the rotating-wave
approximation allows us to neglect all time-dependent terms in the
Hamiltonian, which then takes the form
\begin{equation}
\frac{\hat{H}_{\text{R}}}{\hbar}=\frac{\hat{H}}{\hbar}+g\biggl(\hat{p}\sum_{j=1}^{N}\hat{a}_{j}^{\dagger2}+\mathrm{H.c.}\biggr),
\end{equation}
where $\hat{H}$ is the Hamiltonian (\ref{hamiltonian}) of our model
(with $\varepsilon=2g\alpha_{\mathrm{p}}$). We see that the pump
mode couples to the collective jump operator of our model, $\sum_{j=1}^{N}\hat{a}_{j}^{\dagger2}$.
Hence, adiabatically eliminating the pump under the assumption that
its damping rate $\kappa_{\mathrm{p}}$ is much larger than any other
rate ($\mu$, $J$, $\varepsilon$, $g$, and $\kappa$), it is straightforward
to show \cite{CNB16,CNB-QOnotes} that the reduced state $\hat{\rho}=\mathrm{tr}_{\mathrm{p}}\{\hat{\rho}_{\text{R}}\}$
of the array evolves according to the master equation (\ref{MasterEq})
of our model with $\gamma/N=2g^{2}/\kappa_{\mathrm{p}}$, where $\text{tr}_{\text{p}}$
denotes the partial trace over the pump mode.

As for $\kappa=\gamma_{0}+\Gamma$ in (\ref{MasterEq}), we note that
local decay $\gamma_{0}$ is always present to a certain extent in
superconducting circuits, while incoherent pumping $\Gamma$ can be
implemented in several known ways \cite{CNB14,CNB23,Marthaler11prl,Grajcar08,Astafiev07nat}.

\textcolor{black}{Finally, it is interesting to remark that the direct
coupling of all oscillators to the same four-wave mixer can be avoided
by using a trick similar to that presented in \cite{CNB23}, in which
the normal modes of the array, rather than the physical oscillators,
are used as the model oscillators. The coupling of the mixer to a
single physical oscillator would translate into a coupling to all
model oscillators.}

\section{Conclusions}

In this work, we have shown that incorporating quantum-optical processes
such as pair driving and dissipation into many-body models characteristic
of condensed-matter physics provides access to phenomena beyond the
paradigms traditionally explored by either discipline alone. As a
paradigmatic example, we have considered a bosonic array in which
all bosonic nodes are pair-driven by the same source, leading to collective
rather than local pair dissipation. We have shown that the resulting
model, even in the presence of experimentally unavoidable local linear
dissipation, gives rise to remarkably rich and controllable spatiotemporal
phenomena within the superfluid phase.

In particular, we have proven that the condensates are stabilized
into an exotic configuration in which only the modes belonging to
a closed manifold (Bose surface) in Fourier space are populated, with
the distribution of the population along the manifold being controllable
through the initial conditions or a weak external drive. This allows
us, for example, to stabilize condensates populating a finite number
of $\pm\boldsymbol{k}$ pairs of Fourier wave vectors, which can be
controlled by judiciously choosing a target Bose surface (experimentally
tunable through the detuning $\mu$). This leads to periodic and quasiperiodic
patterns with tunable spatial (quasi-)period and orientation. We have
also explicitly shown how to stabilize a pattern in which all modes
of the closed manifold are equally populated, a state conjectured
to play an important role in open problems of condensed-matter physics
such as high-$T_{\text{c}}$ superconductivity.

In addition, by balancing residual local linear decay through an incoherent
pumping mechanism, we have shown that new constants of motion emerge,
forcing the condensate to acquire nontrivial temporal order in the
form of robust oscillations at the frequencies determined, for each
mode, by the dispersion relation. This behavior generalizes that found
in superfluid time crystals, where particle-number conservation leads
to robust oscillations of the macroscopic wave function at a frequency
set by the chemical potential.

Finally, we have put forward a generic scheme for implementing our
model by exploiting four-wave mixing in superconducting circuits.
This opens the way to the experimental exploration of condensates
with nontrivial spatiotemporal order arising in our unconventional
driven-dissipative model.

\section*{Acknowledgements}

We thank Zi Cai and Germán J. de Valcárcel for their critical reading
of the manuscript and for useful suggestions. C.N.-B. is supported
by the Generalitat Valenciana through project CIDEXG/2023/18 and by
the Ministerio de Ciencia, Innovación y Universidades through project
PID2023-153363NB-C22.

\appendix

\section{From the master equation \\to the GP equations \label{App:MEtoGP}}

In order to derive the GP equations (\ref{GPrealspace}) from the
master equation, we start by noting that the equation of motion of
the expectation value of any operator $\hat{A}$ can be found as
\begin{align}
\partial_{t}\langle\hat{A}\rangle= & \mathrm{tr}\{\hat{A}\partial_{t}\hat{\rho}\}=-\frac{\mathrm{i}}{\hbar}\langle[\hat{A},\hat{H}]\rangle\\
 & +\frac{\gamma}{2N}\left(\langle[\hat{J}^{\dagger},\hat{A}]\hat{J}\rangle+\langle\hat{J}^{\dagger}[\hat{A},\hat{J}]\rangle\right)\nonumber \\
 & +\sum_{\boldsymbol{j}}\gamma_{0}\left(\langle[\hat{a}_{\boldsymbol{j}}^{\dagger},\hat{A}]\hat{a}_{\boldsymbol{j}}\rangle+\langle\hat{a}_{\boldsymbol{j}}^{\dagger}[\hat{A},\hat{a}_{\boldsymbol{j}}]\rangle\right)\nonumber \\
 & +\sum_{\boldsymbol{j}}\Gamma\left(\langle[\hat{a}_{\boldsymbol{j}},\hat{A}]\hat{a}_{\boldsymbol{j}}^{\dagger}\rangle+\langle\hat{a}_{\boldsymbol{j}}[\hat{A},\hat{a}_{\boldsymbol{j}}^{\dagger}]\rangle\right),\nonumber 
\end{align}
where we have used the master equation (\ref{MasterEq}) and defined
$\hat{J}=\sum_{\boldsymbol{j}}\hat{a}_{\boldsymbol{j}}^{\dagger2}$.
Applied to the $\hat{a}_{\boldsymbol{j}}$ operators, we obtain
\begin{align}
\partial_{t}\langle\hat{a}_{\boldsymbol{j}}\rangle= & \mathrm{tr}\{\hat{A}\partial_{t}\hat{\rho}\}=(\mathrm{i}\mu-\gamma_{0}+\Gamma)\langle\hat{a}_{\boldsymbol{j}}\rangle-\mathrm{i}\varepsilon\langle\hat{a}_{\boldsymbol{j}}^{\dagger}\rangle\label{preGPapp}\\
 & +\mathrm{i}\frac{J}{2d}\sum_{\langle\boldsymbol{l}\rangle_{\boldsymbol{j}}}\langle\hat{a}_{\boldsymbol{l}}\rangle-\frac{\gamma}{N}\langle\hat{a}_{\boldsymbol{j}}^{\dagger}\sum_{\boldsymbol{l}}\hat{a}_{\boldsymbol{l}}^{\dagger2}\rangle.\nonumber 
\end{align}
Assume now that the state of the system is coherent at all times,
that is, $\bigotimes_{\boldsymbol{j}}|\psi_{\boldsymbol{j}}(t)\rangle$
with $\hat{a}_{\boldsymbol{j}}|\psi_{\boldsymbol{j}}\rangle=\psi_{\boldsymbol{j}}|\psi_{\boldsymbol{j}}\rangle$
and $\psi_{\boldsymbol{j}}\in\mathbb{C}$. Applying then $\langle\hat{a}_{\boldsymbol{l}_{1}}^{\dagger}...\hat{a}_{\boldsymbol{l}_{m}}^{\dagger}\hat{a}_{\boldsymbol{j}_{1}}...\hat{a}_{\boldsymbol{j}_{n}}\rangle=\psi_{\boldsymbol{l}_{1}}^{*}...\psi_{\boldsymbol{l}_{m}}^{*}\psi_{\boldsymbol{j}_{1}}...\psi_{\boldsymbol{j}_{n}}$
on (\ref{preGPapp}), we obtain
\begin{equation}
\dot{\psi}_{\boldsymbol{j}}=\left(\mathrm{i}\mu-\kappa\right)\psi_{\boldsymbol{j}}-\Bigl(\mathrm{i}\varepsilon+\frac{\gamma}{N}\sum_{\boldsymbol{l}}\psi_{\boldsymbol{l}}^{2}\Bigr)\psi_{\boldsymbol{j}}^{*}+\mathrm{i}\frac{J}{2d}\sum_{\left<\boldsymbol{l}\right>_{\boldsymbol{j}}}\psi_{\boldsymbol{j}}.\label{postGP}
\end{equation}
Finally, by defining normalized variables and parameters $\tilde{t}=Jt$,
$\tilde{\psi}_{\boldsymbol{j}}=\sqrt{\gamma/JN}\psi_{\boldsymbol{j}}$,
$\tilde{\mu}=\mu/J$, $\tilde{\varepsilon}=\varepsilon/J$, and $\tilde{\kappa}=\kappa/J$,
we obtain the GP equations as presented in (\ref{GPrealspace}). Note
however that we removed there the tildes to ease the notation, and
also because the normalization is equivalent to setting $J$ and $\gamma/N$
to one in (\ref{postGP}).

\section{Stationary solutions\protect\label{App:StationarySols}}

Here we discuss in detail the stationary solutions present in the
GP equations (\ref{GPequation}). We start by considering the stationary
equations ($\dot{\alpha}_{\boldsymbol{k}}=0$) for an arbitrary pair
$\pm\boldsymbol{k}$,
\begin{subequations}
\begin{align}
\left(\mathrm{i}\omega_{\boldsymbol{k}}-\kappa\right)\alpha_{\boldsymbol{k}} & =\Bigl(\mathrm{\mathrm{i}}\varepsilon+\sum_{\boldsymbol{q}}\alpha_{\boldsymbol{q}}\alpha_{-\boldsymbol{q}}\Bigr)\alpha_{-\boldsymbol{k}}^{*},\label{statGPapp}\\
\left(\mathrm{i}\omega_{\boldsymbol{k}}+\kappa\right)\alpha_{-\boldsymbol{k}}^{*} & =\Bigl(\mathrm{\mathrm{i}}\varepsilon-\sum_{\boldsymbol{q}}\alpha_{\boldsymbol{q}}^{*}\alpha_{-\boldsymbol{q}}^{*}\Bigr)\alpha_{\boldsymbol{k}}.
\end{align}
\end{subequations}
Taking their product, we obtain
\begin{equation}
\Bigl|\mathrm{i}\varepsilon+\sum_{\boldsymbol{q}}\alpha_{\boldsymbol{q}}\alpha_{-\boldsymbol{q}}\Bigr|^{2}=\omega_{\boldsymbol{k}}^{2}+\kappa^{2}-\varepsilon^{2}.\label{seq}
\end{equation}
Now, the left-hand-side of this expression is independent of $\boldsymbol{k}$,
while the right-hand-side depends on it through $\omega_{\boldsymbol{k}}^{2}$.
Therefore, the excitation of two modes with different absolute value
of the dispersion $|\omega_{\boldsymbol{k}}|$ is incompatible, and
thus, stationary solutions can only populate the modes of Bose levels
at opposite values of the dispersion relation. Note that the Bose
surface, being defined by $\omega_{\boldsymbol{k}}=0$, admits stationary
solutions where it is populated alone. This stationary solutions are
easily found directly from the GP equations as explained in the main
text.

\begin{figure}[t]
\includegraphics[width=1\columnwidth]{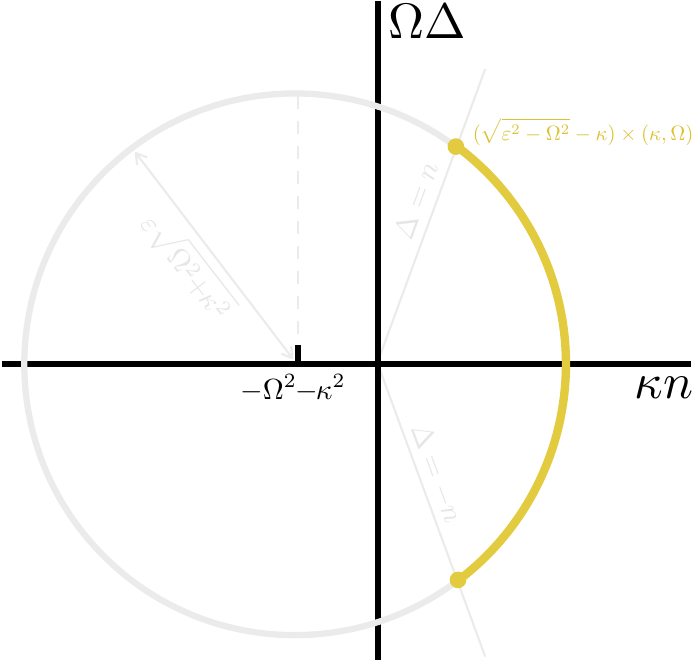} \caption{The stationary configurations outside of the Bose surface can populate
simultaneously two Bose levels $\omega_{\boldsymbol{k}}=\pm\Omega$
with opposite dispersion. $n$ is the total population and $\Delta$
the difference of population between the levels. These stationary
configurations lie on a segment (thick yellow curve) of the circle
shown in the figure in the $(\kappa n,\Omega\Delta)$ space.}
\label{Fig-StationaryCircle} 
\end{figure}

We consider then a stationary solution that populates only two Bose
levels defined by $\omega_{\boldsymbol{k}}=\pm\Omega$, with $\Omega>0$,
which we denote by $\mathrm{BL}_{\pm}$ (the analysis will naturally
accommodate also situations in which one of the $\mathrm{BL}_{\pm}$
does not exist, equivalent to leaving it unpopulated in the upcoming
analysis). In terms of the Bose-level variables (\ref{CollectiveVariables}),
we then have $\sum_{\boldsymbol{q}}\alpha_{\boldsymbol{q}}\alpha_{-\boldsymbol{q}}=s_{+}+s_{-}$.
On the other hand, multiplying Eq. (\ref{statGPapp}) by $\alpha_{-\boldsymbol{k}}$
and summing over all the modes of $\mathrm{BL}_{\pm}$, we obtain
the equations
\begin{subequations}\label{sstatEq}
\begin{align}
(\mathrm{i}\Omega-\kappa)s_{+} & =(\mathrm{i}\varepsilon+s_{+}+s_{-})n_{+},\label{splusStatEq}\\
-(\mathrm{i}\Omega+\kappa)s_{-} & =(\mathrm{i}\varepsilon+s_{+}+s_{-})n_{-}.\label{sminStatEq}
\end{align}
\end{subequations}
Similarly, multiplying Eq. (\ref{statGPapp}) by
$\alpha_{\boldsymbol{k}}^{*}$ and summing over all modes at $\mathrm{BL}_{\pm}$,
we get
\begin{subequations}\label{zstatEq}
\begin{align}
(\mathrm{i}\Omega-\kappa)n_{+} & =(\mathrm{i}\varepsilon+s_{+}+s_{-})s_{+}^{*},\label{zplusStatEq}\\
-(\mathrm{i}\Omega+\kappa)n_{-} & =(\mathrm{i}\varepsilon+s_{+}+s_{-})s_{-}^{*}.\label{zminStatEq}
\end{align}
\end{subequations}
The first thing to note is that $|s_{\pm}|=n_{\pm}$
as follows from (\ref{splusStatEq})/(\ref{zplusStatEq}) and (\ref{sminStatEq})/(\ref{zminStatEq}).
As shown in Appendix \ref{App:LevelConstants} and mentioned in the
text, this implies that the configurations $\alpha_{\boldsymbol{k}}$
at both Bose levels are of the balanced type, which we will also show
here explicitly shortly. Let us now define the total population $n=n_{+}+n_{-}$
and the difference of populations $\Delta=n_{+}-n_{-}$. Operating
on the previous equations as $(\mathrm{i}\Omega-\kappa)($\ref{sminStatEq}$)-(\mathrm{i}\Omega+\kappa)($\ref{splusStatEq}$)$,
it is easy to get
\begin{equation}
\mathrm{i}\varepsilon+s_{+}+s_{-}=\frac{\varepsilon(\Omega^{2}+\kappa^{2})}{\Omega\Delta-\mathrm{i}(\Omega^{2}+\kappa^{2}+\kappa n)}.\label{epsilonss}
\end{equation}
But, at the same time, from the balance condition $|s_{\pm}|=n_{\pm}$
and the absolute value of either (\ref{sstatEq}) or (\ref{zstatEq}),
we know that $|\mathrm{i}\varepsilon+s_{+}+s_{-}|^{2}=\Omega^{2}+\kappa^{2}$,
so equating this to the absolute value of (\ref{epsilonss}) we obtain
\begin{equation}
\Omega^{2}\Delta^{2}+(\Omega^{2}+\kappa^{2}+\kappa n)^{2}=\varepsilon^{2}(\Omega^{2}+\kappa^{2}).\label{DnConstraint}
\end{equation}
We then see that, on the $(\kappa n,\Omega\Delta)$ space, the accessible
configurations lie on a circle of radius $\varepsilon\sqrt{\Omega^{2}+\kappa^{2}}$,
centered at point $-(\Omega^{2}+\kappa^{2},0)$, see Fig. \ref{Fig-StationaryCircle}.
Hence, if we want at least part of the circle to lie on the physical
space $n>0$, the condition $\varepsilon>\sqrt{\Omega^{2}+\kappa^{2}}$
must be satisfied, meaning that the radius is larger than the distance
between the center and the origin of the reference axes. In addition,
the $|\Delta|\leq n$ constraint restricts further the physical region
of the circle (\ref{DnConstraint}), so that only its segment between
points $(\sqrt{\varepsilon^{2}-\Omega^{2}}-\kappa)\times(\kappa,\pm\Omega)$
is available. Note that these points correspond to configurations
where only one Bose level is populated, $\mathrm{BL}_{\pm}$, respectively.
We summarize all these features in Fig. \ref{Fig-StationaryCircle}.
Note also that when $\kappa=0$, the ellipse degenerates into the
lines $|\Delta|=\sqrt{\varepsilon^{2}-\Omega^{2}}$ with $n$ bounded
only $n>|\Delta|$, which is the case we discussed explicitly in the
main text, see Eq. (\ref{BLstatSol}).

Now that we know that $\mathrm{i}\varepsilon+s_{+}+s_{-}=\sqrt{\Omega^{2}+\kappa^{2}}e^{\mathrm{i}\theta}$,
with $\theta=\arg\{\Omega\Delta+\mathrm{i}(\Omega^{2}+\kappa^{2}+\kappa n)\}$,
we can insert it into Eq. (\ref{statGPapp}), which then tells us
that the configuration of the amplitudes at the Bose levels must satisfy
\begin{equation}
\alpha_{\boldsymbol{-k}\in\mathrm{BL}_{\pm}}=-\mathrm{i}e^{\mathrm{i}\varphi_{\pm}}\alpha_{\boldsymbol{k}}^{*},\label{BalanceConditionApp}
\end{equation}
with $\varphi_{\pm}=\arg\left\{ (\pm\Omega+\mathrm{i}\kappa)[\Omega\Delta n+\mathrm{i}(\Omega^{2}+\kappa^{2}+\kappa n)]\right\} $.
Setting $\kappa=0$ we recover the phases $\varphi_{\pm}$ presented
in the main text, Eq. (\ref{BLstatSol}), but now we see that even
when for $\kappa\neq0$, the configurations must be of the balanced
type. Any configuration that satisfies Eqs. (\ref{DnConstraint})
and (\ref{BalanceConditionApp}) is allowed as a stationary solution
of the system, but we show in Appendix \ref{App:BoseLevelStabilityAnalysis}
that they are unstable configurations.

\section{Oscillatory solutions \label{App:OasciSols}}

Our numerics suggest that the oscillatory solutions found for $\kappa=0$
are of the harmonic form $\alpha_{\boldsymbol{k}}(t)=\bar{\alpha}_{\boldsymbol{k}}e^{\mathrm{i}\nu_{\boldsymbol{k}}t}$,
with $\bar{\alpha}_{\boldsymbol{k}}$ independent of time. Inserting
this ansatz into the GP equations (\ref{GPequation}), we obtain
\begin{equation}
(\omega_{\boldsymbol{k}}\negthinspace-\negthinspace\nu_{\boldsymbol{k}})\bar{\alpha}_{\boldsymbol{k}}e^{\mathrm{i}(\nu_{\boldsymbol{k}}+\nu_{-\boldsymbol{k}})t}\negthinspace=\negthinspace\Bigl(\varepsilon-\mathrm{\mathrm{i}}\negthinspace\sum_{\forall\boldsymbol{q}}\bar{\alpha}_{\boldsymbol{q}}\bar{\alpha}_{-\boldsymbol{q}}e^{\mathrm{i}(\nu_{\boldsymbol{q}}+\nu_{-\boldsymbol{q}})t}\Bigr)\bar{\alpha}_{-\boldsymbol{k}}^{*}.\label{OsciEqs}
\end{equation}
In addition, from our exhaustive numerical simulations, we know that
the Bose-level equations always reach asymptotically the fixed point
(\ref{BLeqsFixedPoint}), which means that
\begin{equation}
\sum_{\forall\boldsymbol{k}}\alpha_{\boldsymbol{k}}\alpha_{-\boldsymbol{k}}=\sum_{\beta}s_{\beta}=-\mathrm{i}\varepsilon.
\end{equation}
This is only compatible with the oscillatory solution if $\nu_{-\boldsymbol{k}}=-\nu_{\boldsymbol{k}}$.
The right-hand-side of Eq. (\ref{OsciEqs}) then vanishes, while particularizing
the left-hand-side to a given $\pm\boldsymbol{k}$ pair we obtain
\begin{equation}
(\beta-\nu_{\boldsymbol{k}})\bar{\alpha}_{\boldsymbol{k}}=0=(\beta+\nu_{\boldsymbol{k}})\bar{\alpha}_{-\boldsymbol{k}}.
\end{equation}
For a generic Bose level different than the Bose surface, these conditions
can be satisfied only if one of the amplitudes $\bar{\alpha}_{\pm\boldsymbol{k}}$
vanishes, and the other oscillates with $\nu_{\pm\boldsymbol{k}}=\beta$.
For the Bose surface $\beta=0$, so that these conditions imply $\nu_{\boldsymbol{k}}=0$,
meaning that the asymptotic amplitudes $\alpha_{\boldsymbol{k}}$
are stationary and only constrained by (\ref{BLeqsFixedPoint}).

\section{Interpretation of the level constants \label{App:LevelConstants}}

The level constants (\ref{Cbeta}) can be written as
\begin{equation}
C_{\beta}^{2}=\sum_{\boldsymbol{k}\boldsymbol{k}'\in\mathrm{BL}_{\beta}}\left(|\alpha_{\boldsymbol{k}}|^{2}|\alpha_{\boldsymbol{k}'}|^{2}-\alpha_{\boldsymbol{k}}\alpha_{-\boldsymbol{k}}\alpha_{\boldsymbol{k}'}^{*}\alpha_{-\boldsymbol{k}'}^{*}\right).\label{CbetaApp}
\end{equation}
In the following we show that this constants are positive or zero,
vanishing only when the configuration $\alpha_{\boldsymbol{k}}=\rho_{\boldsymbol{k}}e^{\mathrm{i}\phi_{\boldsymbol{k}}}$
of the system is of the balanced type, that is, $\rho_{\boldsymbol{k}}=\rho_{-\boldsymbol{k}}$
with $\phi_{\boldsymbol{k}}+\phi_{-\boldsymbol{k}}$ fixed to a value
that depends only on the Bose level where $\boldsymbol{k}$ lies in.
In order to see this, we consider two types of terms in the sum (\ref{CbetaApp}),
and bound them separately. First we consider the terms with $\boldsymbol{k}'=\pm\boldsymbol{k}$,
and write explicitly the $\pm\boldsymbol{k}$ contributions to the
sum,
\begin{align}
A_{\boldsymbol{k}}\equiv & |\alpha_{\boldsymbol{k}}|^{4}+|\alpha_{-\boldsymbol{k}}|^{4}+2|\alpha_{\boldsymbol{k}}|^{2}|\alpha_{-\boldsymbol{k}}|^{2}-4\alpha_{\boldsymbol{k}}\alpha_{-\boldsymbol{k}}\alpha_{\boldsymbol{k}}^{*}\alpha_{-\boldsymbol{k}}^{*}.
\end{align}
Next, we consider the terms with $\boldsymbol{k}\neq\boldsymbol{k}'$,
and again write down explicitly the $\pm\boldsymbol{k}$ and $\pm\boldsymbol{k}'$
contributions,
\begin{align}
B_{\boldsymbol{k}\boldsymbol{k}'}\equiv & (|\alpha_{\boldsymbol{k}}|^{2}+|\alpha_{-\boldsymbol{k}}|^{2})(|\alpha_{\boldsymbol{k}'}|^{2}+|\alpha_{-\boldsymbol{k}'}|^{2})\\
 & -2\alpha_{\boldsymbol{k}}\alpha_{-\boldsymbol{k}}\alpha_{\boldsymbol{k}'}^{*}\alpha_{-\boldsymbol{k}'}^{*}-2\alpha_{\boldsymbol{k}'}\alpha_{-\boldsymbol{k}'}\alpha_{\boldsymbol{k}}^{*}\alpha_{-\boldsymbol{k}}^{*}.\nonumber 
\end{align}
In terms of these objects, the level constants can be written as
\begin{equation}
C_{\beta}^{2}=\frac{1}{4}\sum_{\boldsymbol{k}\in\mathrm{BL}_{\beta}}\Bigl(2A_{\boldsymbol{k}}+\underset{\boldsymbol{k}'\neq\pm\boldsymbol{k}}{\sum_{\boldsymbol{k}'\in\mathrm{BL}_{\beta}}}B_{\boldsymbol{k}\boldsymbol{k}'}\Bigr).
\end{equation}
Now we proceed to bound the first type of terms, which is quite trivial
since
\begin{equation}
A_{\boldsymbol{k}}=\rho_{\boldsymbol{k}}^{4}+\rho_{-\boldsymbol{k}}^{4}-2\rho_{\boldsymbol{k}}^{2}\rho_{-\boldsymbol{k}}^{2}=(\rho_{\boldsymbol{k}}^{2}-\rho_{-\boldsymbol{k}}^{2})^{2}\geq0,
\end{equation}
with the equality achieved only if $\rho_{\boldsymbol{k}}=\rho_{-\boldsymbol{k}}$.
The second type of terms requires a bit more work
\begin{align}
B_{\boldsymbol{k}\boldsymbol{k}'}= & \rho_{\boldsymbol{k}}^{2}\rho_{\boldsymbol{k}'}^{2}+\rho_{\boldsymbol{k}}^{2}\rho_{-\boldsymbol{k}'}^{2}+\rho_{-\boldsymbol{k}}^{2}\rho_{\boldsymbol{k}'}^{2}+\rho_{-\boldsymbol{k}}^{2}\rho_{-\boldsymbol{k}'}^{2}\\
 & -2\rho_{\boldsymbol{k}}\rho_{-\boldsymbol{k}}\rho_{\boldsymbol{k}'}\rho_{-\boldsymbol{k}'}\cos(\phi_{\boldsymbol{k}}+\phi_{-\boldsymbol{k}}-\phi_{\boldsymbol{k}'}-\phi_{-\boldsymbol{k}'})\nonumber \\
\geq & (\rho_{\boldsymbol{k}}\rho_{\boldsymbol{k}'}-\rho_{-\boldsymbol{k}}\rho_{-\boldsymbol{k}'})^{2}+(\rho_{\boldsymbol{k}}\rho_{-\boldsymbol{k}'}-\rho_{-\boldsymbol{k}}\rho_{\boldsymbol{k}'})^{2}\geq0,\nonumber 
\end{align}
where in the last line we have upper-bounded the cosine by 1, and
combined the resulting terms into a sum of squares. Note that the
final equality is achieved in this case only when the amplitudes satisfy
the balanced conditions $\phi_{\boldsymbol{k}}+\phi_{-\boldsymbol{k}}=\phi_{\boldsymbol{k}'}+\phi_{-\boldsymbol{k}'}$,
$\rho_{\boldsymbol{k}}=\rho_{-\boldsymbol{k}}$, and $\rho_{\boldsymbol{k}'}=\rho_{-\boldsymbol{k}'}$.
This concludes the proof and shows that $C_{\beta}=0$ if and only
if the configuration of amplitudes $\alpha_{\boldsymbol{k}\in\mathrm{BL}_{\beta}}$
is balanced, otherwise $C_{\beta}>0$.

As a small detail, note that we have considered a nontrivial Bose
level with more than 4 distinct wave vectors. In the one-dimensional
case $d=1$ (where there are at most two wave vectors at a given Bose
level) it's trivial to see that the the level constants vanish when
the amplitudes $\alpha_{\pm k}$ have have equal magnitude, as we
saw in the text, while for single-wave-vector levels with $\boldsymbol{k}=(0,...,0)$
or $-(\pi,...,\pi)$ the level constants are zero by construction.

\section{Stability analysis\protect\label{App:StabilityAnalysis}}

\subsection{Trivial solution\protect\label{App:TrivialStabilityAnalysis}}

The trivial stationary solution reads $\alpha_{\boldsymbol{k}}=0\,\forall\boldsymbol{k}$.
Considering fluctuations $d_{\boldsymbol{k}}$ around it, and expanding
the GP equations (\ref{GPequation}) to first order in these, we obtain
the linear set
\begin{equation}
\left(\begin{array}{c}
\dot{d}_{\boldsymbol{k}}\\
\dot{d}_{-\boldsymbol{k}}^{*}
\end{array}\right)=\left(\begin{array}{cc}
\mathrm{i}\omega_{\boldsymbol{k}}-\kappa & -\mathrm{\mathrm{i}}\varepsilon\\
\mathrm{\mathrm{i}}\varepsilon & -\mathrm{i}\omega_{\boldsymbol{k}}-\kappa
\end{array}\right)\left(\begin{array}{c}
d_{\boldsymbol{k}}\\
d_{-\boldsymbol{k}}^{*}
\end{array}\right).
\end{equation}
The eigenvalues of the linear stability matrix read in this case as
$-\kappa\pm\sqrt{\varepsilon^{2}-\omega_{\boldsymbol{k}}^{2}}$. We
then see that the fluctuations associated to modes $\pm\boldsymbol{k}$
will grow only if the driving satisfies $\varepsilon>\sqrt{\kappa^{2}+\omega_{\boldsymbol{k}}^{2}}$.
Since the Bose-surface modes have $\omega_{\boldsymbol{k}\in\mathrm{BS}}=0$,
these are the modes whose fluctuations get the largest divergence
rate, as long as $\varepsilon>\kappa$. In the $\kappa=0$ case, this
happens for any infinitesimal value of $\varepsilon$.

\subsection{$\mathrm{BS}$ solution\protect\label{App:BSGPstabilityAnalysis}}

Consider now the stationary solution (\ref{BSstatSol}) with only
the Bose-surface modes excited, that is, $\lim_{t\rightarrow\infty}\alpha_{\boldsymbol{k}}(t)=\bar{\alpha}_{\boldsymbol{k}}$,
with $\bar{\alpha}_{\boldsymbol{k}\notin\mathrm{BS}}=0$ and $\bar{\alpha}_{\boldsymbol{k}\in\mathrm{BS}}$
constrained by the condition $\sum_{\boldsymbol{k}\in\mathrm{BS}}\bar{\alpha}_{\boldsymbol{k}}\bar{\alpha}_{-\boldsymbol{k}}=\mathrm{i}(\kappa-\varepsilon)$.
Considering now fluctuations around this solution, that is, $\alpha_{\boldsymbol{k}}(t)=\bar{\alpha}_{\boldsymbol{k}}+d_{\boldsymbol{k}}(t)$,
the GP equations (\ref{GPequation}) are written to first order in
the fluctuations as
\begin{equation}
\dot{d}_{\boldsymbol{k}}=(\mathrm{i}\omega_{\boldsymbol{k}}-\kappa)d_{\boldsymbol{k}}-\mathrm{i}\kappa d_{-\boldsymbol{k}}^{*}-2\bar{\alpha}_{-\boldsymbol{k}}^{*}\sum_{\mathbf{q}\in\mathrm{BS}}\bar{\alpha}_{-\boldsymbol{q}}d_{\boldsymbol{q}}.\label{ddot}
\end{equation}
Let us remark that this equation is valid both for $\kappa$ equal
or different than zero.

Consider first fluctuations off the Bose surface, that is, $\boldsymbol{k}\notin\mathrm{BS}$.
The equations can be recasted as the linear system
\begin{equation}
\left(\begin{array}{c}
\dot{d}_{\boldsymbol{k}}\\
\dot{d}_{-\boldsymbol{k}}^{*}
\end{array}\right)=\left(\begin{array}{cc}
\mathrm{i}\omega_{\boldsymbol{k}}-\kappa & -\mathrm{\mathrm{i}}\kappa\\
\mathrm{\mathrm{i}}\kappa & -\mathrm{i}\omega_{\boldsymbol{k}}-\kappa
\end{array}\right)\left(\begin{array}{c}
d_{\boldsymbol{k}}\\
d_{-\boldsymbol{k}}^{*}
\end{array}\right),
\end{equation}
leading to a linear stability matrix with eigenvalues $-\kappa\pm\sqrt{\kappa^{2}-\omega_{\boldsymbol{k}}^{2}}$.
For $\kappa\neq0$, the real part of these eigenvalues is always negative,
and hence, fluctuations damp back to the stationary solution. In contrast,
when $\kappa=0$ the eigenvalues become $\pm\mathrm{i}\omega_{\boldsymbol{k}}$,
showing that the fluctuations $d_{\pm\boldsymbol{k}}$ remain oscillating
at frequency $\omega_{\boldsymbol{k}}$ around the stationary solution,
without damping or amplification.

The situation is more subtle when the fluctuations lie on the Bose
surface, that is, $\boldsymbol{k}\in\mathrm{BS}$ so that $\omega_{\boldsymbol{k}}=0$.
In this case it is best to consider the evolution equation of the
total fluctuation population, which from (\ref{ddot}) is easily found
to be
\begin{align}
\frac{d}{dt}\left(\sum_{\boldsymbol{k}\in\mathrm{BS}}|d_{\boldsymbol{k}}|^{2}\right)= & -4\left|\sum_{\boldsymbol{k}\in\mathrm{BS}}\bar{\alpha}_{-\boldsymbol{k}}d_{\boldsymbol{k}}\right|^{2}\label{dPopFluc}\\
 & -2\kappa\sum_{\boldsymbol{k}\in\mathrm{BS}}\left(|d_{\boldsymbol{k}}|^{2}+\mathrm{Im}\{d_{\boldsymbol{k}}d_{-\boldsymbol{k}}\}\right).\nonumber 
\end{align}
The first term is obviously negative or zero. It's easy to show that
the second term is also negative or zero, since expressing the fluctuations
in magnitude and phase as $d_{\boldsymbol{k}}=r_{\boldsymbol{k}}e^{\mathrm{i}\varphi_{\boldsymbol{k}}}$,
we have
\begin{align}
\sum_{\boldsymbol{k}\in\mathrm{BS}}\mathrm{Im}\{d_{\boldsymbol{k}}d_{-\boldsymbol{k}}\} & =\sum_{\boldsymbol{k}\in\mathrm{BS}}r_{\boldsymbol{k}}r_{-\boldsymbol{k}}\sin(\varphi_{\boldsymbol{k}}+\varphi_{-\boldsymbol{k}})\\
\geq-\sum_{\boldsymbol{k}\in\mathrm{BS}}r_{\boldsymbol{k}}r_{-\boldsymbol{k}} & \geq-\sum_{\boldsymbol{k}\in\mathrm{BS}}r_{\boldsymbol{k}}^{2}=-\sum_{\boldsymbol{k}\in\mathrm{BS}}|d_{\boldsymbol{k}}|^{2}.\nonumber 
\end{align}
Note that the equality is obtained when $r_{\boldsymbol{k}}=r_{-\boldsymbol{k}}$
and $\varphi_{\boldsymbol{k}}+\varphi_{-\boldsymbol{k}}=-\pi/2\,\forall\boldsymbol{k}$,
which means that the fluctuations perturb the solution in a way compatible
with keeping it balanced, which is a condition that stationary configurations
must satisfy when $\kappa\neq0$, see Eq. (\ref{BSsolLinearDis}).
Note also that when $\kappa=0$ this contribution is directly zero.

This shows that the fate of the fluctuations is either to damp back
to the stationary solution, or remain where they are when all terms
in the right-hand-side of Eq. (\ref{dPopFluc}) are equal to zero.
We have seen that, at least for the second term, this happens when
the fluctuations connect with a new valid stationary solution. In
order to prove that this is the same for the first term, just note
that to first order in the fluctuations
\begin{equation}
\sum_{\boldsymbol{k}\in\mathrm{BS}}\alpha_{\boldsymbol{k}}\alpha_{\boldsymbol{-k}}\approx\sum_{\boldsymbol{k}\in\mathrm{BS}}\bar{\alpha}_{\boldsymbol{k}}\bar{\alpha}_{\boldsymbol{-k}}+2\sum_{\boldsymbol{k}\in\mathrm{BS}}\bar{\alpha}_{-\boldsymbol{k}}d_{\boldsymbol{k}},
\end{equation}
so that fluctuations satisfying $\sum_{\boldsymbol{k}\in\mathrm{BS}}\bar{\alpha}_{-\boldsymbol{k}}d_{\boldsymbol{k}}=0$
lead to configurations that keep the constraint (\ref{BSstatSol})
invariant.

\subsection{$\mathrm{BL}_\beta\neq\mathrm{BS}$ solutions \label{App:BoseLevelStabilityAnalysis}}

The easiest way to show that the stationary solutions at Bose levels
other than the Bose surface are unstable is by performing the linear
stability analysis from the Bose-level equations (\ref{BoseLevelEOM}).
We studied the form of these stationary solutions in Appendix \ref{App:StationarySols},
in particular showing that they satisfy $\mathrm{i}\varepsilon+\sum_{\beta}s_{\beta}=\sqrt{\Omega^{2}+\kappa^{2}}e^{\mathrm{i}\theta}$,
where $\omega_{\boldsymbol{k}}=\pm\Omega$ defines the two Bose levels
that are populated. Remarkably, this is all we need to know in order
to understand the stability of these stationary configurations (and
we don't even need the specific dependence of $\theta$ on the system
parameters). For this, we consider the Bose-level equations (\ref{BoseLevelEOM})
for an unpopulated Bose level $\beta\neq\pm\Omega$, and linearize
them with respect to fluctuations $\delta s_{\beta}(t)$ and $\delta n_{\beta}(t)$
around the stationary solution $s_{\beta}=0=n_{\beta}$. Defining
the vector $\delta\boldsymbol{r}_{\beta}=(\delta s_{\beta},\delta s_{\beta}^{*},\delta n_{\beta})^{T}$,
we obtain a linear system $\delta\dot{\boldsymbol{r}}_{\beta}=\mathcal{L}_{\beta}\delta\boldsymbol{r}_{\beta}$,
with a linear stability matrix (we further define $\sigma=\sqrt{\Omega^{2}+\kappa^{2}}$)
\begin{equation}
\mathcal{L}_{\beta}=-2\left(\begin{array}{ccc}
\kappa-\mathrm{i}\beta & 0 & \sigma e^{\mathrm{i}\theta}\\
0 & \kappa+\mathrm{i}\beta & \sigma e^{-\mathrm{i}\theta}\\
\sigma e^{-\mathrm{i}\theta}/2 & \sigma e^{\mathrm{i}\theta}/2 & \kappa
\end{array}\right),
\end{equation}
which has eigenvalues $-2\kappa$ and $-2\Bigl(\kappa\pm\sqrt{\kappa^{2}+\Omega^{2}-\beta^{2}}\Bigr)$.
The fluctuations $\delta\boldsymbol{r}_{\beta}$ of Bose levels with
$|\beta|<|\Omega|$ will then grow away from the stationary solution,
since one of the eigenvalues is positive for them. This shows that
the stationary solutions at any Bose level other than the Bose surface
(which has the minimum $\beta=0$) are unstable.

\begin{figure*}[t]
\includegraphics[width=0.95\textwidth]{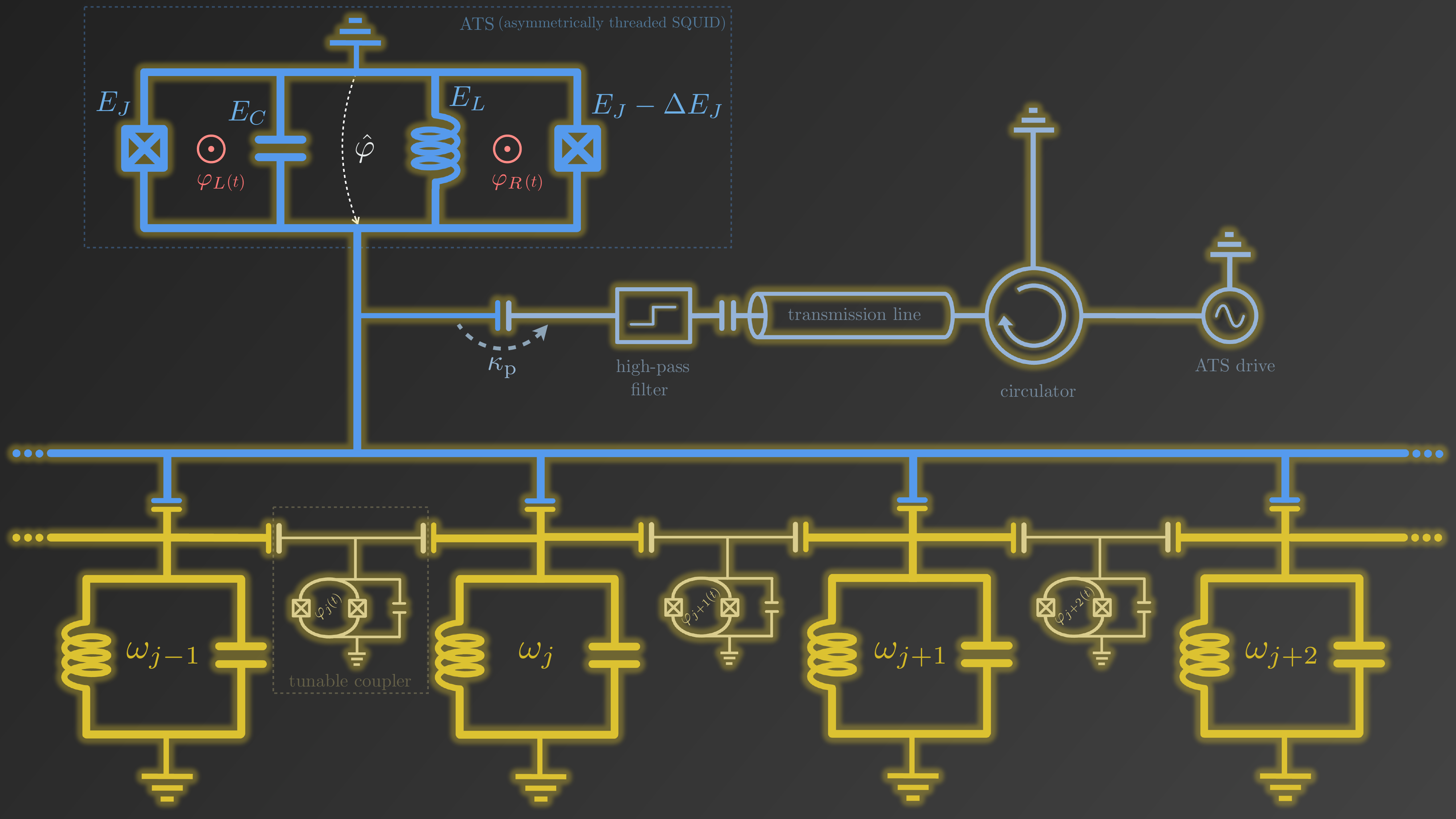} \caption{Sketch of the circuit that allows for the implementation of our model,
based on an asymmetrically-threaded SQUID (blue) coupled to an array
of LC oscillators (yellow) with tunable couplings (light yellow).
The ATS is coupled to a transmission line (light blue) that allows
us to drive it and induces losses at rate $\kappa_{\text{p}}$. A
high-pass filter makes this path invisible to the array.}
\label{Fig-CircuitATS}
\end{figure*}

\section{\textcolor{black}{ATS-based implementation in superconducting circuits}\label{App:ATSimplementation}}

In the main text we show how to implement collective pair driving
and dissipation through a mechanism based on the four-wave mixing
Hamiltonian (\ref{H0_Implementation}) between an array of oscillators,
a lossy driven oscillator of higher frequency, and a classical coherent
polychromatic drive. In this appendix we provide a plausible circuit
implementation of such Hamiltonian based on the asymmetrically threaded
SQUID (ATS) design introduced in \cite{Leghtas20,Leghtas23,Leghtas24}.

We sketch the circuit elements in Fig. \ref{Fig-CircuitATS}. The
ATS (represented in blue) consists of an LC core (with charging and
inductive energies $E_{C}$ and $E_{L}$, respectively) embedded in
a SQUID formed by two junctions (with Josephson energies $E_{J}$
and $E_{J}-\Delta E_{J}$) and threaded by two distinct fluxes $\varphi_{L}(t)$
and $\varphi_{R}(t)$ generated by controllable external magnetic
fields. The ATS is coupled to a transmission line (light blue) from
which we drive it resonantly at rate $\mathcal{E}_{\text{p}}$, also
generating losses at rate $\kappa_{\text{p}}$ (for completeness,
we add also elements such as a filter and a circulator, typically
found along the driving stage in experiments). The ATS is coupled
to the LC resonators that form our array, with distinct frequencies
$\{\omega_{j}\}$, which are in turn interconnected through tunable
couplers, represented in our sketch by transmons whose frequencies
can be modulated with magnetic fluxes $\{\varphi_{j}(t)\}$ (but the
actual coupler design may vary \cite{Campbell23,Yan18,Chen14});
this translates into a time-dependent tunneling rate between neighboring
sites of the array, which we choose as $-J\cos[(\omega_{j}-\omega_{j-1})t]$
between adjacent resonators. According to \cite{Leghtas20,Leghtas23,Leghtas24},
the Hamiltonian describing this design can be written as
\begin{align}
\hat{h}=\hat{h}_{\text{array}}+\hbar\Omega_{\text{p}}\hat{p}^{\dagger}\hat{p}+\mathrm{i}\hbar\mathcal{E}_{\text{p}}\left(e^{-\mathrm{i}\Omega_{\text{p}}t}\hat{p}^{\dagger}-\text{H.c.}\right)+V(\hat{\varphi}),\label{h}
\end{align}
where we have defined the Hamiltonian of the array
\begin{align}
\hat{h}_{\text{array}} & =\sum_{j=1}^{N}\hbar\omega_{j}\hat{a}_{j}^{\dagger}\hat{a}_{j}\\
-\sum_{j=1}^{N} & \hbar J\cos[(\omega_{j}-\omega_{j-1})t]\left(\hat{a}_{j-1}+\hat{a}_{j-1}^{\dagger}\right)\left(\hat{a}_{j}+\hat{a}_{j}^{\dagger}\right),\nonumber 
\end{align}
 and the ATS potential
\begin{align}
V(\hat{\varphi}) & =2\Delta E_{J}\sin(\varphi_{+})\sin(\hat{\varphi}+\varphi_{-})\\
 & \hspace{1em}-2E_{J}\cos(\varphi_{+})\cos(\hat{\varphi}+\varphi_{-}),\nonumber 
\end{align}
with $\varphi_{\pm}(t)=[\varphi_{L}(t)\pm\varphi_{R}(t)]/2$ and a
phase drop across the ATS
\begin{equation}
\hat{\varphi}=\varphi_{\text{p}}\left[\hat{p}+\hat{p}^{\dagger}+\eta\sum_{j}(\hat{a}_{j}+\hat{a}_{j}^{\dagger})\right],
\end{equation}
where $\varphi_{\text{p}}=\left(2E_{C}/E_{L}\right)^{1/4}$ and $\eta$
is the so-called \textit{hybridization} parameter (real and positive)
chosen very small in current designs \cite{Leghtas24}. The only
difference with respect to the design in \cite{Leghtas24} is that,
in our case, we need all circuits of the array to contribute to this
phase drop. However, as mentioned in the main text this condition
could be relaxed by working with the array's eigenmodes, rather than
the physical LC resonators, as the model oscillators \cite{CNB23}.
The operators $\hat{p}$ and $\{\hat{a}_{j}\}$ annihilate excitations
in the ATS circuit (dubbed \textit{pump resonator}) and the circuits
of the array, respectively, and satisfy canonical commutation relations
as mentioned in the main text.

Including the decay of the pump resonator, the state $\hat{\rho}_{\text{i}}$
of this implementation evolves according to the master equation
\begin{equation}
\partial_{t}\hat{\rho}_{\text{i}}=\left[\frac{\hat{h}(t)}{\mathrm{i}\hbar},\hat{\rho}_{\text{i}}\right]+\kappa_{\mathrm{p}}\mathcal{D}_{p}\left[\hat{\rho}_{\text{i}}\right].
\end{equation}

Let us now show how a judicious choice of the circuit parameters turns
this master equation into the one used in the main text, Eq. (\ref{H0}).
Following \cite{Leghtas24}, we choose twin junctions in the ATS
($\Delta E_{J}=0$), and thread with fluxes $\varphi_{L,R}$ such
that
\begin{subequations}
\begin{align}
\varphi_{-} & =\frac{\pi}{2},\\
\varphi_{+} & =\frac{\pi}{2}+\varphi_{d}(t),
\end{align}
\end{subequations}
where
\begin{equation}
\varphi_{\text{d}}(t)=\sum_{m=1}^{N}\bar{\varphi}_{\text{d}}\cos(\nu_{m}t),
\end{equation}
will play the role of the polychromatic driving required in our setting.
As mentioned in the main text, we choose the tones $\nu_{m}=2\omega_{m}-\Omega_{\text{p}}+2\mu$,
where $\mu$ is a small detuning. The ATS potential is then simplified
to
\begin{align}
V(\hat{\varphi}) & =-2E_{J}\sin[\varphi_{\text{d}}(t)]\sin(\hat{\varphi})\label{Vphi_Taylor}\\
 & \approx\frac{E_{J}}{3}\varphi_{\text{d}}(t)\hat{\varphi}^{3}-2E_{J}\varphi_{\text{d}}(t)\hat{\varphi}.\nonumber 
\end{align}
where in the last step we have Taylor-expanded the sine terms around
zero, as commonly done in these systems \cite{Leghtas24} where $\varphi_{\text{d}}$
and $\hat{\varphi}$ are deliberately kept small. The second term
in the potential is linear in the annihilation and creation operators,
and hence acts as a coherent driving on all modes of the problem.
However, under the assumption that $\Omega_{\text{p}}$ and $\{\omega_{j}\}$
are much larger than $E_{J}\bar{\varphi}_{\text{d}}\varphi_{\text{p}}/\hbar$,
this driving can be safely neglected within the rotating-wave approximation,
since all tones $\{\nu_{m}\}$ are far from $\Omega_{\text{p}}$ and
$\{\omega_{j}\}$ by design. On the other hand, the first term in
(\ref{Vphi_Taylor}) provides a quartic nonlinear coupling between
the polychromatic driving, the pump resonator, and the array. Assuming
now that $\Omega_{\text{p}}$ and $\{\omega_{j}\}$ are much larger
than $E_{J}\bar{\varphi}_{\text{d}}\varphi_{\text{p}}^{3}/\hbar$,
we can invoke again the rotating-wave approximation and write
\begin{equation}
\varphi_{\text{d}}(t)\hat{\varphi}^{3}\approx\frac{\bar{\varphi}_{\text{d}}\varphi_{\text{p}}^{3}\eta^{2}}{2}\sum_{mjl=1}^{N}e^{-\mathrm{i}\nu_{m}t}\hat{p}\hat{a}_{j}^{\dagger}\hat{a}_{l}^{\dagger}+\text{H.c.},
\end{equation}
since any the other term is highly off-resonant. We thus see that
the Hamiltonian (\ref{Vphi_Taylor}) takes the form
\begin{align}
\hat{h} & \approx\hat{h}_{\text{array}}+\hbar\Omega_{\text{p}}\hat{p}^{\dagger}\hat{p}+\mathrm{i}\hbar\mathcal{E}_{\text{p}}\left(e^{-\mathrm{i}\Omega_{\text{p}}t}\hat{p}^{\dagger}-\text{H.c.}\right)\\
 & \hspace{1em}+\sum_{mjl=1}^{N}\hbar g\left(e^{-\mathrm{i}\nu_{m}t}\hat{p}\hat{a}_{j}^{\dagger}\hat{a}_{l}^{\dagger}+\text{H.c.}\right),\nonumber 
\end{align}
with $g=E_{J}\bar{\varphi}_{\text{d}}\varphi_{\text{p}}^{3}\eta^{2}/6\hbar$,
and where we can further write
\begin{align}
\hat{h}_{\text{array}} & \approx\sum_{j=1}^{N}\hbar\omega_{j}\hat{a}_{j}^{\dagger}\hat{a}_{j}\\
 & -\sum_{j=1}^{N}\frac{\hbar J}{2}\left(e^{-\mathrm{i}(\omega_{j}-\omega_{j-1})t}\hat{a}_{j-1}\hat{a}_{j}^{\dagger}+\text{H.c.}\right),\nonumber 
\end{align}
by invoking once again the rotating-wave approximation under the assumption
that $\{\omega_{j}\}$ are much larger than $J$. We thus obtain the
model (\ref{H0}) presented in the main text, with the subtlety that
there we are in a displaced picture for the pump, which is more suited
for its adiabatic elimination as well as to see the physics more transparently.
Specifically, let us define the displacement $\hat{D}(t)=e^{\alpha(t)\hat{p}^{\dagger}-\alpha^{*}(t)\hat{p}}$,
with $\alpha(t)$ a time-dependent amplitude that we will choose later.
The state $\hat{\rho}_{0}=\hat{D}^{\dagger}(t)\hat{\rho}_{\text{i}}\hat{D}(t)$
in the displaced picture evolves according to the master equation
\begin{align}
\partial_{t}\hat{\rho}_{0} & =\left[\dot{\alpha}^{*}\hat{p}-\dot{\alpha}\hat{p}^{\dagger}+\frac{\hat{D}^{\dagger}(t)\hat{h}(t)\hat{D}(t)}{\mathrm{i}\hbar},\hat{\rho}_{0}\right]\\
 & \hspace{1em}+\kappa_{\mathrm{p}}\mathcal{D}_{D^{\dagger}(t)pD(t)}\left[\hat{\rho}_{0}\right],\nonumber 
\end{align}
where we have used $\partial_{t}\hat{D}=\hat{D}(t)\left[\dot{\alpha}(\hat{p}^{\dagger}+\alpha^{*})-\text{H.c.}\right]$.
Using next $\hat{D}^{\dagger}(t)\hat{p}\hat{D}(t)=\hat{p}+\alpha$,
this is rewritten as
\begin{align}
\partial_{t}\hat{\rho}_{0} & =\left[\frac{\hat{H}_{0}(t)}{\mathrm{i}\hbar},\hat{\rho}_{0}\right]+\kappa_{\mathrm{p}}\mathcal{D}_{p}\left[\hat{\rho}_{0}\right]\\
 & -\left[\hat{p}^{\dagger}\Bigl(\dot{\alpha}+(\kappa_{\text{p}}+\mathrm{i}\Omega_{\text{p}})\alpha-\mathcal{E}_{\text{p}}e^{-\mathrm{i}\Omega_{\text{p}}t}\Bigr)-\text{H.c.},\hat{\rho}_{0}\right],\nonumber 
\end{align}
where $\hat{H}_{0}$ is the Hamiltonian (\ref{H0_Implementation})
defined in the main text. The first line is precisely the equation
that we provided in the main text, so that choosing a displacement
$\alpha(t)$ that cancels the second line we obtain the desired result.
Specifically, note that the second line vanishes when
\begin{equation}
\alpha(t)=\frac{\mathcal{E}_{\text{p}}}{\kappa_{\text{p}}}e^{-\mathrm{i}\Omega_{\text{p}}t}+\left(\alpha(0)-\frac{\mathcal{E}_{\text{p}}}{\kappa_{\text{p}}}\right)e^{-(\kappa_{\text{p}}+\mathrm{i}\Omega_{\text{p}})t},
\end{equation}
so that, except for extremely short times on the scale of $\kappa_{\text{p}}^{-1}$,
the displacement has the form $\alpha(t)=\alpha_{\text{p}}e^{-\mathrm{i}\Omega_{\text{p}}t}$
that we provided in the main text, with $\alpha_{\text{p}}=\mathcal{E}_{\text{p}}/\kappa_{\text{p}}$.

\bibliography{CollectiveDissBHM}

\end{document}